\documentclass[twocolumn, superscriptaddress, amsmath, secnumarabic, floatfix, amssymb, nobibnotes, aps, pra, reprint]{revtex4-2}
\usepackage{amssymb}
\usepackage{natbib}
\usepackage{graphicx}
\usepackage{amsmath}
\usepackage{wasysym}
\usepackage[bookmarks = false]{hyperref}
\usepackage{xcolor}
\usepackage{lineno}
\usepackage{physics}

\usepackage{nicefrac}
\usepackage{amsfonts}
\usepackage{comment}
\usepackage[utf8]{inputenc}
\begin{document}

\title{Twin-field quantum key distribution without optical frequency dissemination}

\author{Lai Zhou}
\author{Jinping Lin}
\author{Yumang Jing}
\author{Zhiliang Yuan}
\email{yuanzl@baqis.ac.cn}
\affiliation{Beijing Academy of Quantum Information Sciences, Beijing 100193, China}

\begin{abstract}
Twin-field (TF) quantum key distribution (QKD) has rapidly risen as the most viable solution to long-distance secure fibre communication thanks to its fundamentally repeater-like rate-loss scaling.
However, its implementation complexity, if not successfully addressed, could impede or even prevent its advance into real-world. To satisfy its requirement for twin-field coherence, all present setups adopted essentially a gigantic, resource-inefficient interferometer structure that lacks scalability that mature QKD systems provide with simplex quantum links. Here we introduce a novel technique that can stabilise an open channel without using a closed interferometer and has general applicability to phase-sensitive quantum communications. Using locally generated frequency combs to establish mutual coherence, we develop a simple and versatile TF-QKD setup that does not need service fibre and can operate over links of 100~km asymmetry. We confirm the setup’s repeater-like behaviour and obtain a finite-size rate of 0.32~bit/s at a distance of 615.6~km.
\end{abstract}

\maketitle



\noindent\textbf {Introduction.}
Communication at the single photon level enables quantum key distribution (QKD) to achieve a revolutionary milestone in information security, allowing two distant users to establish a cryptographic key with verifiable secrecy\cite{BENNETT14,gisin02_RMP}.
Decades' development has advanced fibre-based QKD systems to a maturity level that is suitable for showcasing long-term, uninterrupted services in real-world networks\cite{Stucki2011,sasaki11,dynes19,Chen21}.
In such systems,
quantum signals experience the loss of the entire link and thus
their maximally achievable rates scale linearly with the channel transmittance ($\eta$) \cite{takeoka14,pirandola17}.
This rate-loss scaling leads to a prohibitively low rate for long haul links, while such links often bear strategic importance for connecting metropolitan cities.
Theoretically, quantum repeaters\cite{Briegel98,duan01} can improve this scaling to $\eta^{\frac{1}{N+1}}$ with $N$ intermediate nodes and thus enable secure communications over arbitrarily long distances,  but require technologies that are  yet to become practical.

Twin-field (TF) QKD protocol \cite{lucamarini18} was recently proposed for practical long-haul quantum communications.
Similarly to the form of QKD that achieves measurement-device-independence\cite{lo12,Braunstein12} (MDI), it uses an intermediate measurement node that halves the signal transmission loss, but extracts information from single photon interference rather than two-photon coincidence so as to gain the advantageous repeater-like rate-loss scaling of $\sqrt{\eta}$.
Security against general attacks was proven for protocol variants \cite{Tamaki18_arXiv,Ma18_PM,Wang18_SNS,Lin18,curty19,Cui19,curras-lorenzo21}, among which sending-not-sending (SNS)\cite{Wang18_SNS} and no-phase-post selection (NPP)\cite{Lin18,curty19,Cui19} remove partially the need for phase slice reconciliation and thus improve 
key generation efficiency.
An exciting collection of experiments\cite{minder19,Wang19,Liu2019_SNS,Zhong19_TF,fang20_502km_PMQKD, Chen2020_509km,pittaluga21,Liu2021_428km,chen21_511km_TFQKD,clivati22,wang22,Chen22_658km_TF-QKD} has successfully validated TF-QKD's superior rate-loss scaling and repeatedly broken the communication distance record, which now stands at 833~km\cite{wang22}.

\begin{figure}[htb]
\includegraphics[width=\columnwidth]{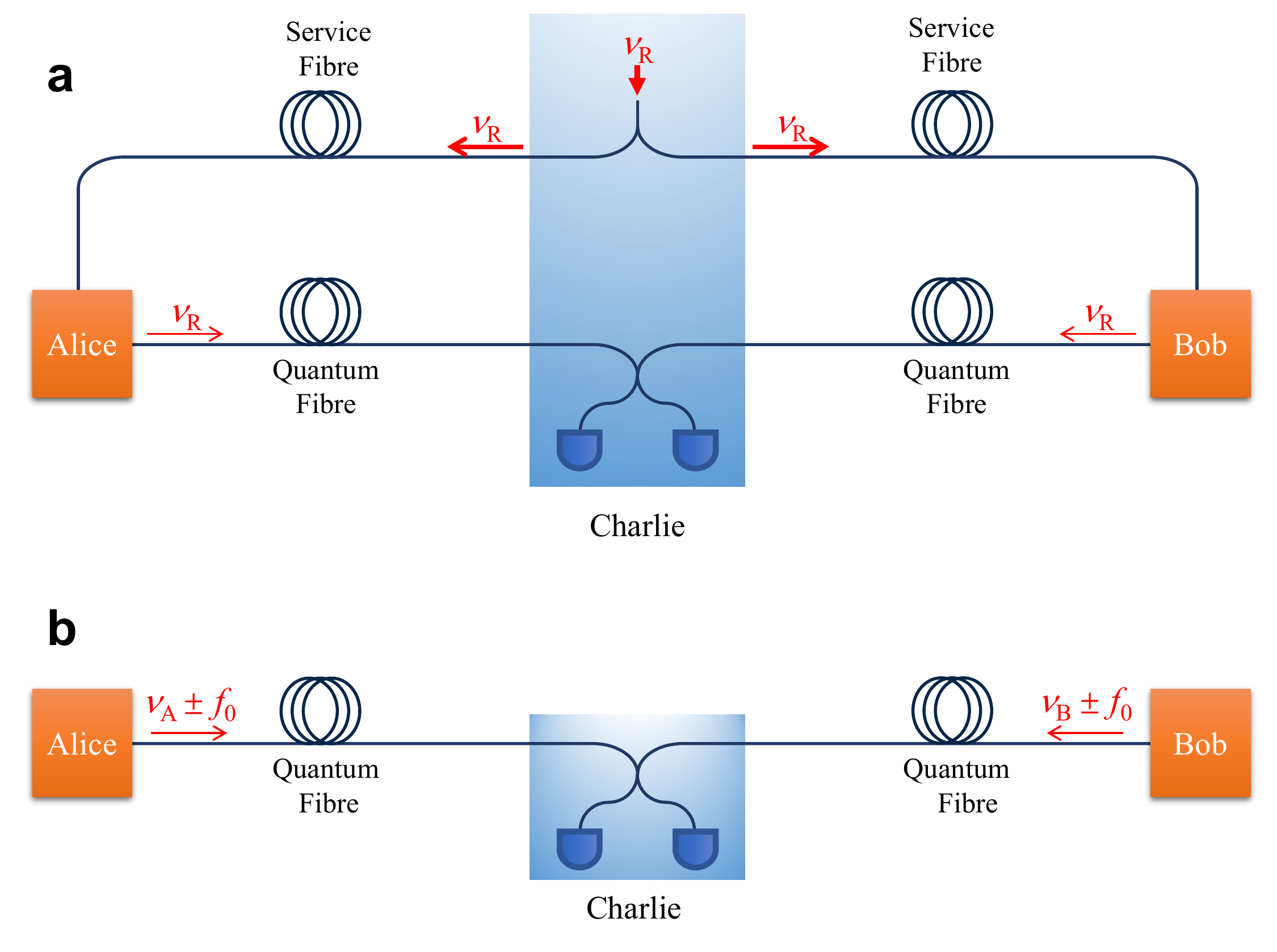}
\caption{\textbf{Schematics of TF-QKD setups.}
In TF-QKD, the users (Alice and Bob) communicate with each other by sending encoded quantum signals at the single-photon level to the intermediate node (Charlie), who measures the interference using two single photon detectors.
The protocol's stringent requirement for phase stability has rendered  all existing setups to adopt a closed interferometer configuration, which is resource-inefficient and inflexible.
\textbf{a}, Existing Mach-Zehnder interferometer setup \cite{pittaluga21,clivati22,wang22}.  Alice and Bob inherit a common optical frequency $\nu_R$ that is disseminated by Charlie via the long service fibres.
\textbf{b}, Open channel setup.  Alice and Bob locally generate their own optical frequencies and their coherent side-bands of $\nu_A \pm f_0$ and $\nu_B \pm f_0$, with a nominally identical microwave frequency offset $f_0$.  One side-band is used to reconcile the laser frequency difference  ($\Delta \nu = \nu_A - \nu_B$), while the other is allocated quantum signal encoding.  The open scheme eliminates the need for the service fibre and the optical frequency locking hardware, and supports asymmetric links.
}
\label{fig:concept}
\end{figure}

With above achievements, the primary experimental focus should now turn to addressing TF-QKD's implementation complexity, which would otherwise impair its practical deployment.  Due to the stringent requirement for twin-field phase tracking,  all existing long-haul TF-QKD setups have to adopt essentially a gigantic Mach-Zehnder inteferometer (MZI) configuration with half of its fibre used for optical frequency dissemination (Fig.~\ref{fig:concept}\textbf{a}).  Use of service fibre brings two severe drawbacks.
First,  the setups are fibre-resource inefficient and require additional frequency locking hardware and often optical amplifiers along the service fibre.
Second,  the closed fibre configuration is inherently incompatible with optical switching, which could restrict TF-QKD's scalability into a larger network like other QKD systems \cite{toliver03,tang16_MDI_Network}.
Using Sagnac interferometer permits a simple TF-QKD ring network \cite{zhong22_TF_Network}, but its long-haul capability could be obstructed by noise contamination due to counter-propagating signals of strong intensity disparity.
Ideally, the twin-field phase could be stabilised without using a closed interferometer so as to reach a simple setup (Fig.~\ref{fig:concept}\textbf{b}) sharing an identical fibre configuration as MDI-QKD.
We note that new MDI-QKD variants were recently proposed to allow repeater-like rate-loss scaling via post-detection time-bin pairing\cite{Xie22_AsynchronousMDI-QKD,Zeng22_MP}.

\noindent{\textbf{Open quantum link stabilisation.}}
To perform TF-QKD, it is necessary to ensure stable interference at the intermediate node (Charlie) between signals transmitted by two remote users (Alice and Bob).
The differential phase ($\phi$) between their signals evolves as
\begin{equation}
    \dv{\phi}{t} = 2\pi(\Delta \nu + \frac{\nu}{s} \dv{\Delta L}{t}),
\label{eq:drift}
\end{equation}
\noindent where $\Delta \nu$ is the difference between the users' laser frequencies ($\nu$), $s$ light speed in the fibre, and $\Delta L$ the length difference between the users' fibres to Charlie.
The fibre term ($\frac{\nu}{s}\dv{\Delta L}{t}$) contributes typically a few kHz to the phase instability for a fibre link of several hundred kilometers\cite{lucamarini18,pittaluga21}.
This kHz drift alone can be corrected for
with a feedback loop of a sub-MHz bandwidth using a low-level reference signal, without scattering overwhelming noise photons and thus crucially maintaining the quantum channel intact.
However, the laser term is more problematic.
Free-running lasers have unsatisfying long-term stability, with daily frequency drift often in the region of 10 - 100~MHz, although some can offer instantaneous linewidth of 1~kHz.  Referencing to a high-fineness cavity helps, but the cavity itself drifts. Consequently, TF-QKD setups to date resort to sending strong laser signals to synchronise the users' lasers to enforce $\Delta \nu = 0$ and hence require a separate service fibre channel to avoid contamination to the quantum link (see Fig.~\ref{fig:concept}\textbf{a}).  The resulting setup has a closed fibre configuration.

In this work, we develop a scheme to stabilise an open quantum link between two distant users that share no prior mutual coherence and are separated by hundreds kilometers of single mode fibre.
Using this scheme we have achieved a drastically simplified and versatile TF-QKD setup, capable of supporting asymmetric  links, that needs neither dissemination of optical frequencies nor its associated service fibre and hardware, as schematically shown in Fig.~\ref{fig:concept}\textbf{b}.
Here, each user generates their local frequency comb and transmits one comb line to Charlie for interference, the outcome of which feeds into a photon-counting proportional–integral–derivative (PID) controller that allows rapid reading out and zeroing the laser frequency difference as well as cancelling the fibre fluctuation.
Importantly, our solution brings neither performance degradation nor loss of practicality thanks partly to its choice of proven ingredients including ultra-stable lasers\cite{chen21_511km_TFQKD,Chen22_658km_TF-QKD}, optical frequency combs\cite{clivati22} and dual-wavelength stabilisation \cite{pittaluga21,clivati22}.

\begin{figure*}[ht]
\includegraphics[width=1.8\columnwidth]{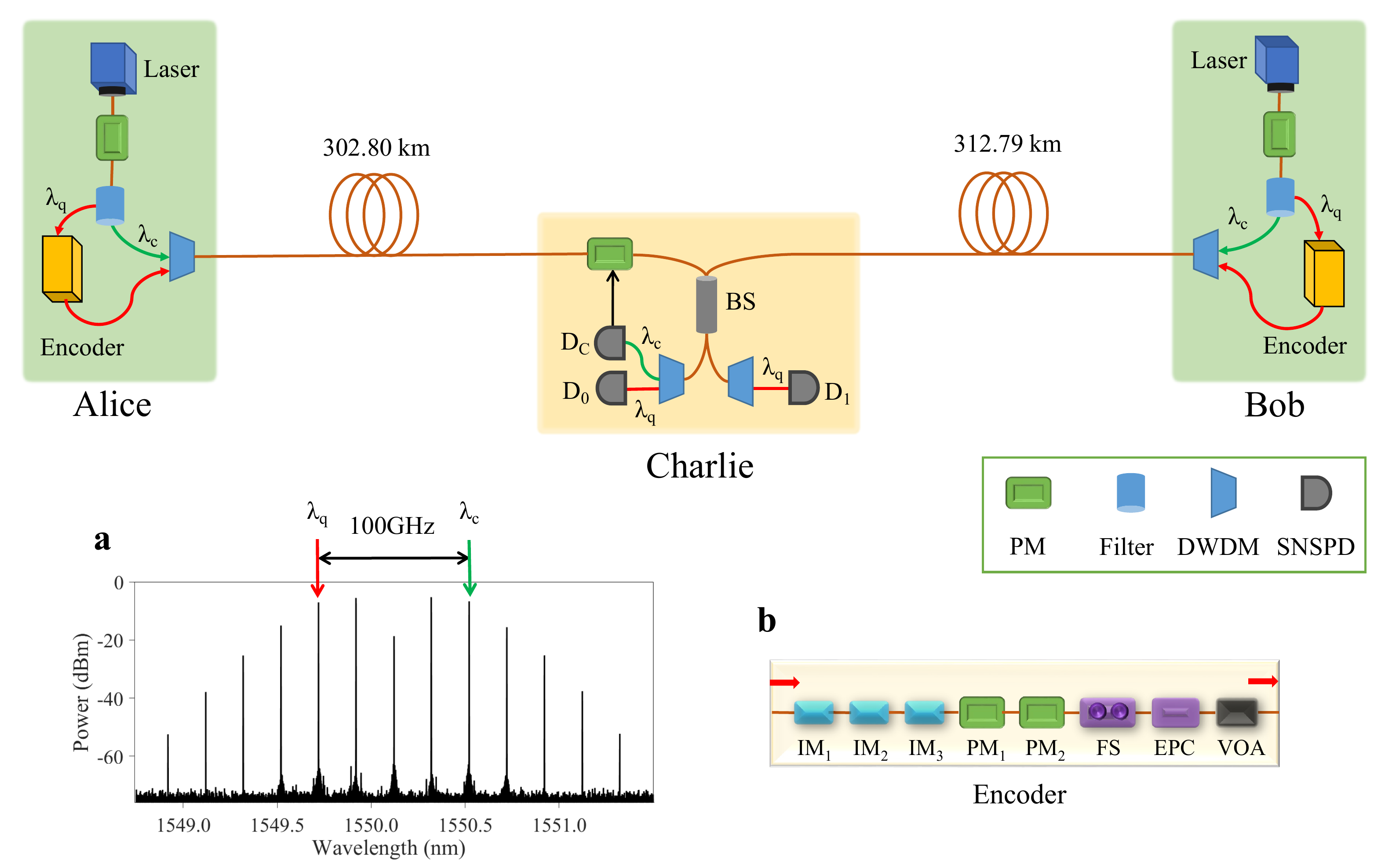}
\caption{\textbf{Experimental setup.}   Alice (Bob) owns an independent ultrastable laser, the signal of which is modulated by a phase modulator (PM) to produce a frequency comb of 25~GHz spacing.  Two comb lines separated by  100~GHz are chosen for quantum signal encoding  ($\lambda_q$)
and  channel stabilisation  ($\lambda_c$), respectively.  Charlie contains a receiving 50/50 beam splitter to interfere the incoming signals.  The $\lambda_q$ photons are registered by D$_0$ and D$_1$ and the $\lambda_c$ photons by D$_c$.  D$_c$'s count rate is used as error signal to the fast PID controller that cancels the twin-field phase fluctuation of the $\lambda_c$ signals. \textbf{a}, Optical frequency comb measured after the PM; \textbf{b}, Encoder box.  DWDM: dense wavelength-division multiplexing; EPC: electrically driven polarisation controller; FS: fibre stretcher; IM: Intensity Modulator;  VOA: variable optical attenuator.}
\label{fig:Setup}
\end{figure*}

\noindent\textbf{Setup.} Our setup (Fig.~\ref{fig:Setup}) does not need service fibre and has an identical fibre configuration as MDI-QKD \cite{lo12}.
Alice and Bob are connected to Charlie from opposite directions via their segments of the quantum link. The quantum link is made of spools of ultra-low loss fibres with an average loss coefficient of 0.168~dB km$^{-1}$. For detailed information on the fibre properties, refer to Supplementary Table~S1.

Each user owns an independent continuous-wave laser that features a sub-Hz short-term linewidth and allows adjustment of its optical frequency at 1~mHz step size.  See Methods for the detailed information on the lasers.
Passing through a phase modulator (PM), the laser light is modulated to generate a frequency comb (Fig.~\ref{fig:Setup}\textbf{a}) with a precise spacing of 25~GHz thanks to the microwave frequency driver referenced to a Rubidium frequency standard.
For 50~GHz dense-wavelength-division-multiplexing (DWDM) compatibility, two comb lines of $\lambda_q=1549.72$~nm and $\lambda_c=1550.52$~nm (ITU Channels 34.5 and 33.5) are selected and filtered into two separate paths. The continuous-wave $\lambda_c$ signal is used to establish coherence with the other user, and we refer to it as the `channel reference'.
The $\lambda_q$ signal passes through the encoder box (Fig.~\ref{fig:Setup}\textbf{b}), which carves the continuous-wave input into a train of pulses of 300~ps width at an interval of 1~ns.   The encoder's description is provided in Supplementary.
We apply blockwise modulation for every 200~pulses. The first 95 pulses of each block do not receive further modulation and are used to sense the phase of the quantum channel. They are referred to as the `quantum reference'.  The last 100 pulses are modulated according to TF-QKD protocol's requirement. As these quantum pulses can be considerably weaker than the quantum reference, the encoder extinguishes the 5 pulses in between to create an empty buffer to prevent inter-group contamination.  Overall, our setup has an effective QKD clock rate of 500~MHz.
After the encoder box, the $\lambda_q$ light is recombined with the channel reference ($\lambda_c$) into the quantum channel segment using a DWDM filter.
In addition to setting the approximate photon fluxes, both wavelengths paths have separate polarisation controls for pre-compensation of polarisation rotation by the quantum link.

\begin{figure*}[ht]
\includegraphics[width=1.9\columnwidth]{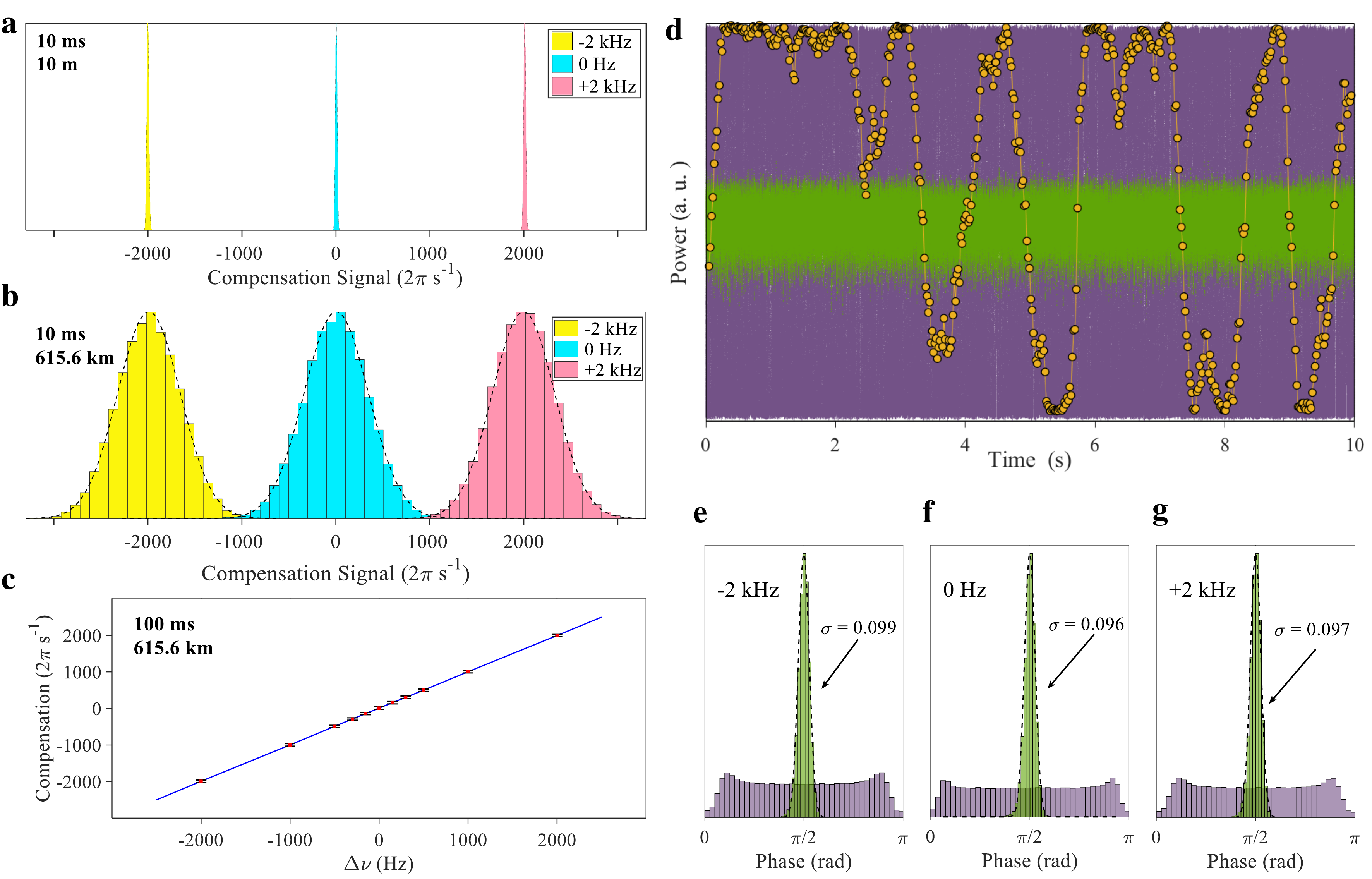}
\caption{\textbf{Open quantum channel stabilisation.} Alice and Bob's lasers are fully independent, and their frequency difference is adjustable by offsetting Alice's laser to a high-fineness cavity. Except \textbf{a}, all data in this figure were measured with the 615.6~km quantum fibre. \textbf{a}, Histograms of phase compensation signals integrated over 10~ms time intervals for a 10~m quantum link; \textbf{b}, same as \textbf{a} but with 615.6~km quantum link; \textbf{c},
Compensation signal angular frequency as a function of the laser frequency offset ($\Delta \nu$).
\textbf{d},  Optical output power as a function of time measured at one output of Charlie's 50/50 interfering beam-splitter; Purple (green): channel reference or the $\lambda_c$ signal when the FPGA PID controller is turned off (on); Orange: slowed drift of the $\lambda_q$ signal. \textbf{e}, \textbf{f}, \textbf{g}, The phase angle distributions of the channel reference  before (purple) and after (green) the simplex channel stabilisation, measured for laser frequency offsets of -2, 0 and 2~kHz, respectively.   Data in \textbf{d} -- \textbf{g} were measured with a power meter.
}
\label{fig:Stablization}
\end{figure*}

After travelling through their respective quantum link segments, Alice and Bob's light enters Charlie's 50/50 interfering beam splitter with an identical polarisation and matched intensities.
The interference outcome is spectrally de-multiplexed before detection by three superconducting nanowire single photon detectors (SNSPD's), with D$_0$ and D$_1$ assigned to the $\lambda_q$ signals and D$_{c}$ for the $\lambda_c$ channel reference. Charlie contains a PM in one of his input arms for fast phase feedback control.  Charlie's components losses and detector performance is summarised in Supplementary Tables S2 and S3.

We describe briefly below, and provide more information in Section II of Supplementary, on how our setup is stabilised.  For all different fibre lengths,  we adjust the channel references' intensities to have a maximal interference visibility and maintain an average count rate of approximately 13~MHz at D$_{c}$.  This count rate allows 200~kHz sampling with acceptable noise by a field-programmable-gated-array (FPGA) PID controller (not shown) to process and generate compensation voltages to Charlie's PM, locking the differential phase to $\pi/2$ between the channel references.
This locking process cancels simultaneously both instability terms in Eq. (1), \textit{i.e.}, the laser frequency difference and fibre fluctuation.
As described later, the FPGA controller allows real-time readout of the laser frequency difference and can thus apply feedback to one user's laser to prevent the laser drifting out of the PID's correction bandwidth.
Due to coherence among comb lines,  the frequency difference for the $\lambda_q$ signals is reduced by a factor of $\frac{\abs{\lambda_q - \lambda_c}}{\lambda_q}$.
The phase drift by this residual difference can then be corrected for through a second stage compensation \cite{pittaluga21} which uses the interference result of the quantum reference to act on a fibre stretcher in Alice's encoder at a rate of 50--100~Hz.
Note that use of coherent combs enables the setup to support asymmetric channels, which can substantially ease fibre provision during installation.

\noindent\textbf{Experimental results.}
To evaluate the effectiveness of our open fibre scheme, we analyse the compensation signal that is applied to the Charlie's PM in response to the lasers' frequency difference ($\Delta \nu$) and fibre fluctuation of the quantum channel.
Figs.~\ref{fig:Stablization}\textbf{a} and \ref{fig:Stablization}\textbf{b} show the histograms of compensation signals integrated over 10~ms intervals for three different frequency offsets.
For clarity, the PM compensation signal is converted to angular frequency with unit of $2\pi$~s$^{-1}$ or Hz. For a 10~m quantum channel (Fig.~\ref{fig:Stablization}\textbf{a}),  the PID compensates mostly the laser frequency difference and we therefore observe a histogram of sharp distributions with its compensating signal tracking exactly the laser frequency difference.
After installing 615.6~km fibre (Fig.~\ref{fig:Stablization}\textbf{b}), each histogram becomes considerably broadened because the PID has now to work on also rapid and random fibre fluctuation.
However, the peak of each histogram remains at the same angular frequency, as the random fibre fluctuation does not produce a net drift. We can then determine the laser frequency drift with high accuracy by increasing the readout time to just  100~ms, as shown in Fig.~\ref{fig:Stablization}\textbf{c}, where the compensation rate follows strictly the frequency offset.
This fast readout enables real-time compensation for laser frequency drift, which is indispensable for long-term operation of TF-QKD.

Fig.~\ref{fig:Stablization}\textbf{d} shows the interference results of the channel references at Charlie's interfering 50/50 beam-splitter after travelling the 615.6~km quantum channel. With the PID switched off, the measured optical power oscillates violently between constructive and destructive extremes due to fast fibre fluctuation.  We extract an average fringe visibility of 97.8~\%, which is noticeably worse than the value of 99.0~\% obtained with a single laser serving both the users, illustrating a small penalty of using truly independent lasers.
We extract a standard deviation of 1.07~kHz for the fibre drift, which is in good agreement with data extracted from the PID control (not shown) and  comparable to those reported in the literature\cite{lucamarini18,chen21_511km_TFQKD,pittaluga21,wang22}.
After enabling the PID control,
the interference output is narrowed down to a tight band (green).  We extract the phase angles and plot their distributions as shown in Figs.~\ref{fig:Stablization}\textbf{e} -- \ref{fig:Stablization}\textbf{g}.
We obtain almost identical standard deviations of 0.099, 0.096 and 0.097~rad for different frequency offsets of -2, 0 and +2~kHz, suggesting the PID control is tolerant to at least 2~kHz detuning.
A phase deviation of $0.1$~rad will cause about 0.5~\% drop in the interference visibility, which is acceptable for TF-QKD.
After the PID control on the channel reference,  the phase drift of the $\lambda_q$ signals is drastically slowed down, as shown by orange circles in Fig.~\ref{fig:Stablization}\textbf{d}.
The standard deviation of this drift rate is 0.72~Hz, which is 1500 times slower than the value for the unstabilised channel.  We measure an interference fringe visibility of be 96.8~\%, which will cause 1.6~\% floor to the QBER in the phase basis. For detailed information on how the visibility values were extracted, refer to Section~IV in Supplementary.

With the simplex channel stabilisation, we performed two sets of TF-QKD experiments  using the SNS protocol\cite{Wang18_SNS,xu20_SNS_AOPP_original,Jiang20_zigzag_AOPP}.
In the first set, the users' channel losses to Charlie are strictly matched while their fibre lengths may differ by 10~km, which was introduced for compensating the loss by Charlie's PM.
We ran the SNS protocol for three distances of $403.73$, $518.16$ and $615.59$ km, with $2.025 \times 10^{12}$, $2.475 \times 10^{12}$ and $1.418 \times 10^{13}$ total pulses sent respectively.
We take the actively odd-parity pairing (AOPP) method for data post-processing, which can efficiently reduce the bit-flip error rate of the raw key and have a higher probability for pairing success than the random `two-way classical communication' method \cite{xu20_SNS_AOPP_original}, and use the zigzag approach proposed in Ref. \cite{Jiang20_zigzag_AOPP} in the secure key rate (SKR) calculation.
The detailed experimental parameters and results can be found in Supplementary Tables S4-S6.
In Fig.~\ref{fig:SKR}\textbf{a}, we present our experimental results (red square) in terms of SKR versus distance together with the simulation curve (red line).
We include also the absolute repeaterless Pirandola-Laurenza-Ottaviani-Banchi bound \cite{pirandola17} (black line), \textit{i.e.}, SKC$_0$,
which represents the fundamentally maximum rate that an ideal point-point QKD could achieve and can only be overcome by a repeater-like setup.
With finite-size effects being taken into consideration,
we obtain SKR's of 146.7, 14.38 and 0.32~bit/s for 403.73, 518.16 and 615.59~km, respectively.
They all overcome the SKC$_0$ bound, confirming the repeater-like behavior of our setup. At 615.6~km, the SKR is 9.70 times above SKC$_0$.

\begin{figure}[ht]
\includegraphics[width=1.0\columnwidth]{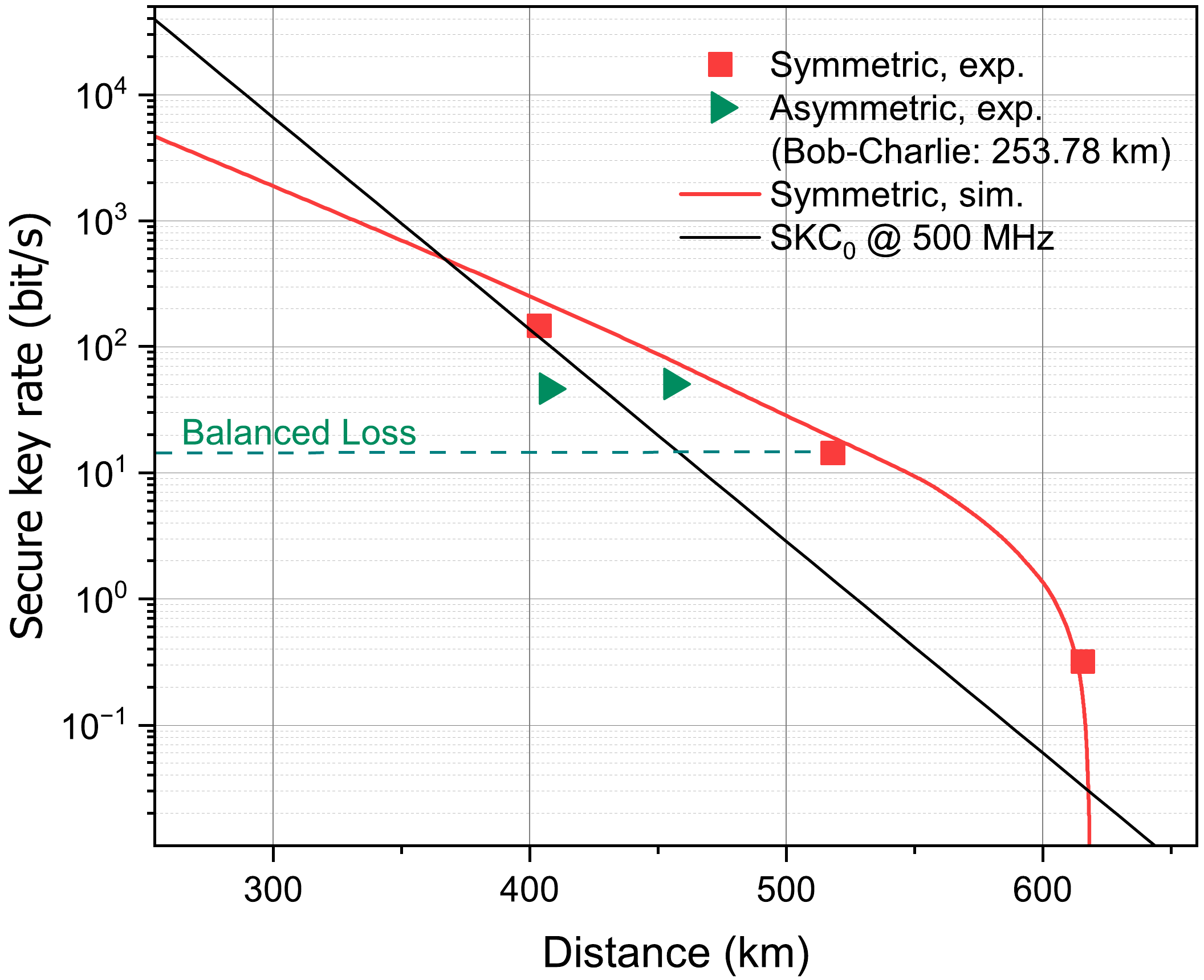}
\caption{\textbf{Secure key rate (SKR) simulations and results.}
The AOPP-SNS TF-QKD with finite size effects was implemented in the experiments.
Two sets of experimental data are included. Symmetric case (squares):  The users' losses to Charlie are strictly matched while their fibre lengths may differ by 10~km; Asymmetric case (triangles): Bob's fibre length is fixed at 253.78~km.
The green dashed line indicates the expected SKR when the asymmetry to Bob's 253.78~km fibre is treated by just adding attenuation.
A fibre attenuation coefficient of 0.168~dB~km$^{-1}$ is adopted for calculating the SKR simulation (red line) and the absolute repeaterless SKC$_0$ bound (black line) for an ideal point-point QKD setup operating at 500~MHz.}
\label{fig:SKR}
\end{figure}

To further demonstrate the robustness of our open channel scheme, we explore the capability of our setup supporting asymmetric fibre links.
With Bob's fibre fixed at $253.78$~km,  we ran experiments over two different distances of $201.87$ and $153.45$~km between Alice and Charlie, representing a link asymmetry of $51.91$ and $100.33$~km, respectively. For fair assessment, we use an identical parameter set and compensate the extra loss asymmetry by adding $8.13$~dB attenuation to the $153.45$~km link.
During optimisation of the parameter set, we apply mathematical constraint that need to be satisfied for the security of asymmetric SNS protocol\cite{hu19_asymmetric_SNS} and introduce a new constraint for the intensities of both decoy states to guarantee a high interference visibility at Charlie.  The experimental results are shown in Fig.~\ref{fig:SKR} (green triangles).
We extract respective finite-size SKR's of $50.75$ and $46.30$~bit/s for the asymmetries of $51.91$~km and $100.33$~km,  with a minor deterioration by extra 48~km asymmetry.  This result improves considerably over the current asymmetry record of 22~km \cite{clivati22}, while most setups had to keep channels strictly matched down to meters \cite{minder19,Wang19,pittaluga21,wang22}.
Additionally, our result illustrates also the importance of parameter optimization for asymmetric links, as the obtained SKR's are thrice higher than the rate (green dashed line) expected if we just add attenuation to balance the fibre disparity.

\begin{table*}[ht]
\centering
\caption{A selection of recent state-of-the-art TF-QKD experiments and comparison with this work.  All previous long-haul setups use optical frequency dissemination technologies, including
homodyne/heterodyne optical phase locked loop (OPLL)\cite{minder19,Wang19,pittaluga21,clivati22,wang22},
time-frequency metrology\cite{Liu2019_SNS,Chen2020_509km,Chen22_658km_TF-QKD} and optical injection locking \cite{fang20_502km_PMQKD,Liu2021_428km},
to synchronise the users' lasers with the reference optical frequency delivered via a service fibre that is as long as the quantum fibre (see Fig.~\ref{fig:concept}\textbf{a}).  The subsequent twin-field phase can be either actively stabilised \cite{pittaluga21,clivati22,Wang19,wang22} or reconciled through  post-selection at the data processing stage \cite{fang20_502km_PMQKD,chen21_511km_TFQKD,Chen22_658km_TF-QKD}, though the latter approach does not support NPP-TF-QKD protocols.  Among active schemes, using a second wavelength \cite{pittaluga21,clivati22} can suppress double Rayleigh scattering noise and increase the stabilisation bandwidth, while further introduction of inter-wavelength  coherence\cite{clivati22} enables support for asymmetric fibre channels.  The X-basis quantum bit error rate (QBER) is listed just for experiments that adopted the SNS protocol for fair comparison.
}
\label{my-label}
\begin{tabular}{c|c|c|c|c|c|c|c}
\hline \hline
    Experiment & \begin{tabular}[c]{@{}c@{}}Quantum\\Fibre\end{tabular} &
    \begin{tabular}[c]{@{}c@{}}Service\\Fibre\end{tabular} &
    \begin{tabular}[c]{@{}c@{}}Frequency\\ Dissemination\end{tabular}& \begin{tabular}[c]{@{}c@{}}Phase \\ Compen- \\sation\end{tabular} &   \begin{tabular}[c]{@{}c@{}} Number of \\ wavelengths \end{tabular} &        \begin{tabular}[c]{@{}c@{}} Inter- \\wavelength\\Coherence\end{tabular} &
    \begin{tabular}[c]{@{}c@{}} X-Basis\\QBER\end{tabular}
    \\ \hline 
\hline
\begin{tabular}[c]{@{}c@{}} Chen \textit{et al.}\cite{Chen22_658km_TF-QKD},\\ 2022 \end{tabular}
& 658.7~km & 500~km &  \begin{tabular}[c]{@{}c@{}}Time-freq. \\ metrology\end{tabular} & Passive & 1 & n/a & $\sim$5.0 \%\\
\hline
\begin{tabular}[c]{@{}c@{}}Wang \textit{et al.}\cite{wang22},\\2022 \end{tabular}&  833.8~km&  833.8~km  &  \begin{tabular}[c]{@{}c@{}}Homodyne\\  OPLL \end{tabular}
& Active  &  1 & n/a  & n/a \\
\hline
\begin{tabular}[c]{@{}c@{}} Pittaluga \textit{et al.}\cite{pittaluga21},\\ 2021 \end{tabular} &    605.2~km &
611.4~km &  \begin{tabular}[c]{@{}c@{}}Heterodyne\\ OPLL\end{tabular}
& Active & 2 & No & 5.41 \%\\
\hline
\begin{tabular}[c]{@{}c@{}} Clivati \textit{et al.}\cite{clivati22},\\ 2022 \end{tabular} &       206~km &
206~km & \begin{tabular}[c]{@{}c@{}}Heterodyne\\ OPLL \end{tabular}  & Active  &  2 & Yes & n/a \\
\hline
\begin{tabular}[c]{@{}c@{}} This work\end{tabular}
&615.6~km & Not needed & Not needed & Active & 2 & Yes & 4.75 \%\\
\hline \hline
\end{tabular}
\label{table:recent_experiments}
\end{table*}

We compare our open scheme with recent experiments that adopted the closed MZI configuration.
As SKR's and distances are directly affected by the detector performance and/or the clock frequency, the relevant parameter to compare here is the quantum bit error rate (QBER) arising from the interfering twin-field  signals that went through TF-QKD's phase randomisation process.
Based on this criteria,  we list in Table~\ref{table:recent_experiments} the X-basis QBER for experiments that adopted also the SNS protocol.
Within the margin of error,  our system gives even a slightly better QBER of 4.75~\% than the other setups \cite{pittaluga21,Chen22_658km_TF-QKD}, showing no performance degradation from using truly independent lasers between an open quantum channel.

\noindent\textbf{Discussion.} We have shown for the first time the capability of stabilising the phase of a simplex quantum link without resorting to a closed interferometer-type configuration.
With the open-channel stabilisation technique we are able to achieve a simple and robust TF-QKD setup that can drastically ease fibre provision and route planning for future deployment.  The technique could be adapted to enable free-space quantum experiments involving single-photon interference between remote light sources, including free-space TF-QKD.
Moreover, its demonstrated frequency tolerance makes it possible to use less-demanding lasers, \textit{e.g.}, these lasers\cite{Maurice20,Talvard17} that reference to absolute atomic or molecular transitions and feature typically 1~kHz linewidth, stimulating further simplification in TF-QKD setups.
We believe our technique is applicable to phase-based quantum applications in general, including quantum repeaters \cite{duan01}, single photon entanglement distribution\cite{Caspar20} and quantum internet \cite{Pompili21}.


\subsection*{Methods}

\noindent\textbf{Ultra-stable lasers.} The ultra-stable lasers were manufactured by MenloSystems (Model: ORS-Cubic). Each laser emits at a wavelength of 1550.12~nm and features a sub-Hz short-term linewidth thanks to its use of Pound-Drever-Hall (PDH) technique for locking to a cavity with a fineness of 250000 and  a free spectral range (FSR) of 3~GHz.
In the PDH locking path there is an extra phase modulator which is driven at a base frequency of 300~MHz and allows fine adjustment of the laser frequency with a step size of 1~mHz.
We measured the two lasers to have respective frequency drift rates of 110.4~mHz~s$^{-1}$ and 92.5~mHz~s$^{-1}$ in the same direction, and their differential frequency drifts about 1500~Hz per day.
During initial characterisation of the TF-QKD setup, the frequency difference ($\Delta \nu$) between the lasers was monitored through their beating note recorded by a photodiode and can be precisely set via adjusting the modulation frequency to the PM in one laser.  This laser frequency drift can easily be corrected for by our photon-counting-based FPGA PID controller, so the offset monitoring by the photodiode is unnecessary during TF-QKD experiments. Our laser frequency feedback will work even when they are installed in separated locations.

\noindent\textbf{Coherent frequency comb.} We generate the electro-optic frequency comb by passing the continuous-wave laser through a phase modulator that is driven by a stable microwave source with its power carefully set.
Driving at 25~GHz, we obtain a total of 13 comb lines between 1548.9~nm to 1551.3~nm as shown in Fig.~\ref{fig:Setup}\textbf{a}.  The comb lines of 1549.72~nm ($\lambda_q$) and 1550.52~nm ($\lambda_c$) are selected as the quantum signal and the channel reference respectively.  Their spacing of 0.8~nm (100~GHz) compatible with the ITU G694.1 standard DWDM grid and allows convenient spectral filtering and wavelength routing.
After spectral filtering, each comb line has an output power of $\geq$300~$\mu$W and channel isolation of $>$55~dB, both sufficient for TF-QKD encoding.

\noindent\textbf{Protocol.} In this experiment, we adopt a 4-intensity SNS-TF-QKD protocol with actively odd-parity pairing (AOPP)~\cite{xu20_SNS_AOPP_original} for the data post-processing, with finite-size effects being taken into account. We describe the theory of asymmetric protocol \cite{hu19_asymmetric_SNS} and mention that it also applies to the symmetric case when Alice and Bob have an identical loss to Charlie and use the same values for their source parameters.

In this protocol, Alice and Bob repeat the following process $N_{\text{tot}}$ times to obtain a string of binary bits. In each time window, Alice (Bob) randomly decides whether it is a decoy window with probability $p_{Ax}$ ($p_{Bx}$) or a signal window with probability $1-p_{Ax}$ ($1-p_{Bx}$). If it is a signal window, Alice (Bob) randomly prepares a phase-randomized weak coherent state (WCS) with intensity $\mu_{Az}$ ($\mu_{Bz}$) and decides whether to send it or not, with probabilities $\epsilon_A$ ($\epsilon_B$) and $1-\epsilon_A$ ($1-\epsilon_B$), respectively. For the decisions of not-sending in the signal windows, Alice (Bob) denotes them as bit 0 (1), and for the decisions of sending, Alice (Bob) denotes them as bit 1 (0). If a decoy window is chosen, Alice (Bob) randomly prepares phase-randomized WCS with intensities $\mu_{A0}$ ($\mu_{B0}$), $\mu_{A1}$ ($\mu_{B1}$) or $\mu_{A2}$ ($\mu_{B2}$) with respective probabilities $1-p_{A1}-p_{A2}$ ($1-p_{B1}-p_{B2}$), $p_{A1}$ ($p_{B1}$) and $p_{A2}$ ($p_{B2}$).  As was proven in Ref. \cite{hu19_asymmetric_SNS}, in order to main the security of the protocol, the asymmetric source parameters should satisfy the following mathematical constraint
%
\begin{align}
    \frac{\mu_{A1}}{\mu_{B1}} = \frac{\epsilon_{A} (1-\epsilon_{B} )\mu_{Az} e^{-\mu_{Az}}}{\epsilon_{B} (1-\epsilon_{A} )\mu_{Bz} e^{-\mu_{Bz}}}.
\label{eq:constraint}
\end{align}
%
Note that this requirement is automatically met for the symmetric protocol, and its purpose is to guarantee the users' raw key free from systematic bias. After the preparation stage, Alice and Bob send their pulses to the middle untrusted node, Charlie, who performs interference measurements and announces publicly which detector clicks. The detection events which one and only one detector clicks are taken as effective events. For time windows determined by both Alice and Bob to be a signal window, which are labelled as Z windows, Alice and Bob get two $n_t$ bits of raw key strings comprised by the corresponding bits from effective events. The bit-flip error rate of these two strings is denoted as $E_z$.

Events in time windows determined by both Alice and Bob to be a decoy window, which are labelled as X windows, are used to perform the security analysis. In X windows where Alice and Bob choose intensities $\mu_{A1}$ and $\mu_{B1}$, respectively, the phase information of their WCSs would be publicly announced and post-selected based on the following criteria
\begin{widetext}
\begin{align}
    | \theta_{A1} - \theta_{B1} | \leq \frac{2 \pi}{M} \quad\quad\quad \text{or} \quad\quad\quad | \theta_{A1} - \theta_{B1}-\pi | \leq \frac{2 \pi}{M},
\end{align}
\end{widetext}
where $\theta_{A1}$ and $\theta_{B1}$ are the private phases of Alice's and Bob's pulses respectively and $M$ is the number of phase slices.

The AOPP method is used to reduce the errors of raw key strings before the error correction and privacy amplification processes. In AOPP, Bob first actively pairs his bits 0 with bits 1 of the raw key string and announces the pairing information to Alice. Alice performs the same pairing accordingly and they then compare the parity of pairs. They discard both bits in the pair if the announced parities are different and keep the first bit of the pairs if the parities are the same. The users now use the remaining bits to form a new shorter string $n_t'$ with dramatically reduced bit-flip error rate $E_z'$, from which they will extract the final key.

Next we briefly show how to extract the information of single-photon states, which constitutes the final key rate formula, from X-window events, through decoy-state analysis. We denote the counting rate of sources $\kappa\zeta$ in X windows by $S_{\kappa\zeta}$, which is the ratio of the number of corresponding effective events to the number of respective pulses sent out by Alice and Bob. These values can be measured in the experiment. Note that a statistical fluctuation analysis should be considered here as part of the finite-size effects, with more details being presented in Ref.~\cite{Yu19_SNS_statistical}. Then we use the decoy-state method to deduce the counting rate of single-photon states which either Alice or Bob actually sends out a single photon from WCSs, which are \cite{hu19_asymmetric_SNS},
\begin{widetext}
\begin{align}
     \langle \underline{y_{10}} \rangle &= \frac{\mu_{A2}^2 e^{\mu_{A1}} \langle \underline{S}_{\mu_{A1}\mu_{B0}} \rangle - \mu_{A1}^2 e^{\mu_{A2}} \langle \overline{S}_{\mu_{A2}\mu_{B0}} \rangle - (\mu_{A2}^2-\mu_{A1}^2) \langle \overline{S}_{\mu_{A0}\mu_{B0}} \rangle}{\mu_{A2} \mu_{A1} (\mu_{A2}-\mu_{A1})},\\
    \langle \underline{y_{01}} \rangle &= \frac{\mu_{B2}^2 e^{\mu_{B1}} \langle \underline{S}_{\mu_{A0}\mu_{B1}} \rangle - \mu_{B1}^2 e^{\mu_{B2}} \langle \overline{S}_{\mu_{A0}\mu_{B2}} \rangle - (\mu_{B2}^2-\mu_{B1}^2) \langle \overline{S}_{\mu_{A0}\mu_{B0}} \rangle}{\mu_{B2} \mu_{B1} (\mu_{B2}-\mu_{B1})},
\end{align}
\end{widetext}
%
respectively, where the notations $\langle \underline{\cdot} \rangle$ and $\langle \overline{\cdot} \rangle$ denote the lower and the upper bound of the corresponding expected values, respectively, with a composable definition of security and the Chernoff bound being applied. Their detailed explanations and expressions can be found in the Ref. \cite{Jiang19_SNS_composable_security}.
Then the lower bound of the expected value of the counting rate of untagged bits, which is the number of bits generated through effective events when the users actually send out single-photon states in Z windows, is given by \cite{hu19_asymmetric_SNS}
%
\begin{align}
    \langle \underline{y_{1}} \rangle = \frac{\mu_{A1}}{\mu_{A1}+\mu_{B1}}  \langle\underline{y_{10}} \rangle + \frac{\mu_{B1}}{\mu_{A1}+\mu_{B1}} \langle \underline{y_{01}} \rangle
\end{align}
%
and the lower bound of the expected value of the number of untagged bits is
\begin{widetext}
\begin{align}
   \langle\underline{n_{1}} \rangle = N_{\text{ZZ}} \left[ \epsilon_{A} (1-\epsilon_{B}) \mu_{Az} e^{-\mu_{Az}} \langle \underline{y_{10}} \rangle + \epsilon_{B} (1-\epsilon_{A}) \mu_{Bz} e^{-\mu_{Bz}} \langle \underline{y_{01}} \rangle \right],
\end{align}
\end{widetext}
%
%
%
where $N_{\text{ZZ}}=N_{\text{tot}}(1-p_{Ax})(1-p_{Bx})$ is the number of pulses Alice and Bob both choose signal windows. According to Ref. \cite{hu19_asymmetric_SNS}, so long as Eq.~\ref{eq:constraint} is satisfied, the phase-flip error rate of untagged bits in Z windows can be calculated from the bit-flip error rate of untagged bits in X windows, which is written as
%
\begin{align}
  \langle\overline{e_1^{ph}}\rangle  = \frac{\langle\overline{T}_{XX}\rangle - 1/2 e^{-\mu_{A1}-\mu_{B1}} \langle \underline{S}_{\mu_{A0}\mu_{B0}} \rangle}{e^{-\mu_{A1}-\mu_{B1}} (\mu_{A1}+\mu_{B1}) \langle \underline{y_1} \rangle}
\end{align}
%
where $T_{XX}$ is the ratio of the number of corresponding error events over the number of total pulses with intensities $\mu_{A1}$ and $\mu_{B1}$ sent out in X windows.

The secret key rate (SKR) of AOPP with finite-size effects is given by
%
\begin{align}
    R=\frac{1}{N_{\text{tot}}} \{ n_1' [1-h(e_1'^{\text{ph}})] - f n_t' h(E_z') - \Delta \},
\end{align}
%
where $h (x) =-x \text{log}_2 x - (1-x) \text{log}_2 (1-x)$ is the binary Shannon entropy. $f=1.1$ is the error correction efficiency factor. $\Delta=2 \text{log}_2 (2/\epsilon_{\text{cor}}) + 4 \text{log}_2 (1/\sqrt{2}\epsilon_{\text{PA}} \hat{\epsilon})$ is the finite-size correction term, with $\epsilon_{\text{cor}} {\color{black} =10^{-10}}$, $\epsilon_{\text{PA}}{\color{black} =10^{-10}}$ and $\hat{\epsilon}{\color{black} =10^{-10}}$ being the failure probabilities for error correction, privacy amplification and the coefficient of the smoothing parameter, respectively. $n_1'$ and $e_1'^{\text{ph}}$ are the number of untagged bits and their phase-flip error rate, respectively, after AOPP process. Here we adopt a zigzag approach proposed in Ref. \cite{Jiang20_zigzag_AOPP} in order to obtain higher key rates and take all the finite-key effects efficiently, the same as the calculation method applied in experimental reference~\cite{Liu2021_428km,chen21_511km_TFQKD,Chen22_658km_TF-QKD}. For simplicity, we do not list the calculation processes here, with all details can be found in the cited papers.

\subsection*{Acknowledgement}

We thank X. B. Wang for helpful discussions on the SNS-AOPP TF-QKD protocol.
Z. L. acknowledges support by the National Natural Science Foundation of China (Grant number 62105034).


\begin{thebibliography}{10}
\expandafter\ifx\csname url\endcsname\relax
  \def\url#1{\texttt{#1}}\fi
\expandafter\ifx\csname urlprefix\endcsname\relax\def\urlprefix{URL }\fi
\providecommand{\bibinfo}[2]{#2}
\providecommand{\eprint}[2][]{\url{#2}}

\bibitem{BENNETT14}
\bibinfo{author}{Bennett, C.~H.} \& \bibinfo{author}{Brassard, G.}
\newblock \bibinfo{title}{Quantum cryptography: Public key distribution and
  coin tossing}.
\newblock \emph{\bibinfo{journal}{Theor. Comput. Sci.}}
  \textbf{\bibinfo{volume}{560}}, \bibinfo{pages}{7--11}
  (\bibinfo{year}{2014}).

\bibitem{gisin02_RMP}
\bibinfo{author}{Gisin, N.}, \bibinfo{author}{Ribordy, G.},
  \bibinfo{author}{Tittel, W.} \& \bibinfo{author}{Zbinden, H.}
\newblock \bibinfo{title}{Quantum cryptography}.
\newblock \emph{\bibinfo{journal}{Rev. Mod. Phys.}}
  \textbf{\bibinfo{volume}{74}}, \bibinfo{pages}{145--195}
  (\bibinfo{year}{2002}).

\bibitem{Stucki2011}
\bibinfo{author}{Stucki, D.} \emph{et~al.}
\newblock \bibinfo{title}{Long-term performance of the {SwissQuantum} quantum
  key distribution network in a field environment}.
\newblock \emph{\bibinfo{journal}{New J. Phys.}} \textbf{\bibinfo{volume}{13}},
  \bibinfo{pages}{123001} (\bibinfo{year}{2011}).

\bibitem{sasaki11}
\bibinfo{author}{Sasaki, M.} \emph{et~al.}
\newblock \bibinfo{title}{Field test of quantum key distribution in the tokyo
  qkd network}.
\newblock \emph{\bibinfo{journal}{Opt. Express}} \textbf{\bibinfo{volume}{19}},
  \bibinfo{pages}{10387--10409} (\bibinfo{year}{2011}).

\bibitem{dynes19}
\bibinfo{author}{Dynes, J.~F.} \emph{et~al.}
\newblock \bibinfo{title}{Cambridge quantum network}.
\newblock \emph{\bibinfo{journal}{npj Quant. Inf.}}
  \textbf{\bibinfo{volume}{5}}, \bibinfo{pages}{101} (\bibinfo{year}{2019}).

\bibitem{Chen21}
\bibinfo{author}{Chen, Y.-A.} \emph{et~al.}
\newblock \bibinfo{title}{An integrated space-to-ground quantum communication
  network over 4,600 kilometres}.
\newblock \emph{\bibinfo{journal}{Nature}} \textbf{\bibinfo{volume}{589}},
  \bibinfo{pages}{214 -- 219} (\bibinfo{year}{2021}).

\bibitem{takeoka14}
\bibinfo{author}{Takeoka, M.}, \bibinfo{author}{Guha, S.} \&
  \bibinfo{author}{Wilde, M.~M.}
\newblock \bibinfo{title}{Fundamental rate-loss tradeoff for optical quantum
  key distribution}.
\newblock \emph{\bibinfo{journal}{Nat. Commun.}} \textbf{\bibinfo{volume}{5}},
  \bibinfo{pages}{5235} (\bibinfo{year}{2014}).

\bibitem{pirandola17}
\bibinfo{author}{Pirandola, S.}, \bibinfo{author}{Laurenza, R.},
  \bibinfo{author}{Ottaviani, C.} \& \bibinfo{author}{Banchi, L.}
\newblock \bibinfo{title}{Fundamental limits of repeaterless quantum
  communications}.
\newblock \emph{\bibinfo{journal}{Nat. Commun.}} \textbf{\bibinfo{volume}{8}},
  \bibinfo{pages}{15043} (\bibinfo{year}{2017}).

\bibitem{Briegel98}
\bibinfo{author}{Briegel, H.-J.}, \bibinfo{author}{D\"ur, W.},
  \bibinfo{author}{Cirac, J.~I.} \& \bibinfo{author}{Zoller, P.}
\newblock \bibinfo{title}{Quantum repeaters: The role of imperfect local
  operations in quantum communication}.
\newblock \emph{\bibinfo{journal}{Phys. Rev. Lett.}}
  \textbf{\bibinfo{volume}{81}}, \bibinfo{pages}{5932--5935}
  (\bibinfo{year}{1998}).

\bibitem{duan01}
\bibinfo{author}{Duan, L.-M.}, \bibinfo{author}{Lukin, M.~D.},
  \bibinfo{author}{Cirac, J.~I.} \& \bibinfo{author}{Zoller, P.}
\newblock \bibinfo{title}{Long-distance quantum communication with atomic
  ensembles and linear optics}.
\newblock \emph{\bibinfo{journal}{Nature}} \textbf{\bibinfo{volume}{414}},
  \bibinfo{pages}{413 -- 418} (\bibinfo{year}{2001}).

\bibitem{lucamarini18}
\bibinfo{author}{Lucamarini, M.}, \bibinfo{author}{Yuan, Z.~L.},
  \bibinfo{author}{Dynes, J.~F.} \& \bibinfo{author}{Shields, A.~J.}
\newblock \bibinfo{title}{Overcoming the rate-distance limit of quantum key
  distribution without quantum repeaters}.
\newblock \emph{\bibinfo{journal}{Nature}} \textbf{\bibinfo{volume}{557}},
  \bibinfo{pages}{400--403} (\bibinfo{year}{2018}).

\bibitem{lo12}
\bibinfo{author}{Lo, H.-K.}, \bibinfo{author}{Curty, M.} \&
  \bibinfo{author}{Qi, B.}
\newblock \bibinfo{title}{Measurement-device-independent quantum key
  distribution}.
\newblock \emph{\bibinfo{journal}{Phys. Rev. Lett.}}
  \textbf{\bibinfo{volume}{108}}, \bibinfo{pages}{130503}
  (\bibinfo{year}{2012}).

\bibitem{Braunstein12}
\bibinfo{author}{Braunstein, S.~L.} \& \bibinfo{author}{Pirandola, S.}
\newblock \bibinfo{title}{Side-channel-free quantum key distribution}.
\newblock \emph{\bibinfo{journal}{Phys. Rev. Lett.}}
  \textbf{\bibinfo{volume}{108}}, \bibinfo{pages}{130502}
  (\bibinfo{year}{2012}).

\bibitem{Tamaki18_arXiv}
\bibinfo{author}{Tamaki, K.}, \bibinfo{author}{Lo, H.-K.},
  \bibinfo{author}{Wang, W.} \& \bibinfo{author}{Lucamarini, M.}
\newblock \bibinfo{title}{Information theoretic security of quantum key
  distribution overcoming the repeaterless secret key capacity bound}
  (\bibinfo{year}{2018}).
\newblock \bibinfo{note}{ArXiv: 1805.05511}.

\bibitem{Ma18_PM}
\bibinfo{author}{Ma, X.}, \bibinfo{author}{Zeng, P.} \& \bibinfo{author}{Zhou,
  H.}
\newblock \bibinfo{title}{Phase-matching quantum key distribution}.
\newblock \emph{\bibinfo{journal}{Phys. Rev. X}} \textbf{\bibinfo{volume}{8}},
  \bibinfo{pages}{031043} (\bibinfo{year}{2018}).

\bibitem{Wang18_SNS}
\bibinfo{author}{Wang, X.-B.}, \bibinfo{author}{Yu, Z.-W.} \&
  \bibinfo{author}{Hu, X.-L.}
\newblock \bibinfo{title}{Twin-field quantum key distribution with large
  misalignment error}.
\newblock \emph{\bibinfo{journal}{Phys. Rev. A}} \textbf{\bibinfo{volume}{98}},
  \bibinfo{pages}{062323} (\bibinfo{year}{2018}).

\bibitem{Lin18}
\bibinfo{author}{Lin, J.} \& \bibinfo{author}{L\"utkenhaus, N.}
\newblock \bibinfo{title}{Simple security analysis of phase-matching
  measurement-device-independent quantum key distribution}.
\newblock \emph{\bibinfo{journal}{Phys. Rev. A}} \textbf{\bibinfo{volume}{98}},
  \bibinfo{pages}{042332} (\bibinfo{year}{2018}).

\bibitem{curty19}
\bibinfo{author}{Curty, M.}, \bibinfo{author}{Azuma, K.} \&
  \bibinfo{author}{Lo, H.-K.}
\newblock \bibinfo{title}{Simple security proof of twin-field type quantum key
  distribution protocol}.
\newblock \emph{\bibinfo{journal}{npj Quant. Inf.}}
  \textbf{\bibinfo{volume}{5}}, \bibinfo{pages}{64} (\bibinfo{year}{2019}).

\bibitem{Cui19}
\bibinfo{author}{Cui, C.} \emph{et~al.}
\newblock \bibinfo{title}{Twin-field quantum key distribution without phase
  postselection}.
\newblock \emph{\bibinfo{journal}{Phys. Rev. Appl.}}
  \textbf{\bibinfo{volume}{11}}, \bibinfo{pages}{034053}
  (\bibinfo{year}{2019}).

\bibitem{curras-lorenzo21}
\bibinfo{author}{Curr\'{a}s-Lorenzo, G.} \emph{et~al.}
\newblock \bibinfo{title}{Tight finite-key security for twin-field quantum key
  distribution}.
\newblock \emph{\bibinfo{journal}{npj Quant. Inf.}}
  \textbf{\bibinfo{volume}{7}}, \bibinfo{pages}{22} (\bibinfo{year}{2021}).

\bibitem{minder19}
\bibinfo{author}{Minder, M.} \emph{et~al.}
\newblock \bibinfo{title}{Experimental quantum key distribution beyond the
  repeaterless rate-loss limit}.
\newblock \emph{\bibinfo{journal}{Nat. Photonics}}
  \textbf{\bibinfo{volume}{13}}, \bibinfo{pages}{334--338}
  (\bibinfo{year}{2019}).

\bibitem{Wang19}
\bibinfo{author}{Wang, S.} \emph{et~al.}
\newblock \bibinfo{title}{Beating the fundamental rate-distance limit in a
  proof-of-principle quantum key distribution system}.
\newblock \emph{\bibinfo{journal}{Phys. Rev. X}} \textbf{\bibinfo{volume}{9}},
  \bibinfo{pages}{021046} (\bibinfo{year}{2019}).

\bibitem{Liu2019_SNS}
\bibinfo{author}{Liu, Y.} \emph{et~al.}
\newblock \bibinfo{title}{Experimental twin-field quantum key distribution
  through sending or not sending}.
\newblock \emph{\bibinfo{journal}{Phys. Rev. Lett.}}
  \textbf{\bibinfo{volume}{123}}, \bibinfo{pages}{100505}
  (\bibinfo{year}{2019}).

\bibitem{Zhong19_TF}
\bibinfo{author}{Zhong, X.}, \bibinfo{author}{Hu, J.}, \bibinfo{author}{Curty,
  M.}, \bibinfo{author}{Qian, L.} \& \bibinfo{author}{Lo, H.-K.}
\newblock \bibinfo{title}{Proof-of-principle experimental demonstration of
  twin-field type quantum key distribution}.
\newblock \emph{\bibinfo{journal}{Phys. Rev. Lett.}}
  \textbf{\bibinfo{volume}{123}}, \bibinfo{pages}{100506}
  (\bibinfo{year}{2019}).

\bibitem{fang20_502km_PMQKD}
\bibinfo{author}{Fang, X.-T.} \emph{et~al.}
\newblock \bibinfo{title}{Implementation of quantum key distribution surpassing
  the linear rate-transmittance bound}.
\newblock \emph{\bibinfo{journal}{Nat. Photonics}}
  \textbf{\bibinfo{volume}{14}}, \bibinfo{pages}{422--425}
  (\bibinfo{year}{2020}).

\bibitem{Chen2020_509km}
\bibinfo{author}{Chen, J.-P.} \emph{et~al.}
\newblock \bibinfo{title}{Sending-or-not-sending with independent lasers:
  Secure twin-field quantum key distribution over 509 km}.
\newblock \emph{\bibinfo{journal}{Phys. Rev. Lett.}}
  \textbf{\bibinfo{volume}{124}}, \bibinfo{pages}{070501}
  (\bibinfo{year}{2020}).

\bibitem{pittaluga21}
\bibinfo{author}{Pittaluga, M.} \emph{et~al.}
\newblock \bibinfo{title}{600-km repeater-like quantum communications with
  dual-band stabilization}.
\newblock \emph{\bibinfo{journal}{Nat. Photonics}}
  \textbf{\bibinfo{volume}{15}}, \bibinfo{pages}{530--535}
  (\bibinfo{year}{2021}).

\bibitem{Liu2021_428km}
\bibinfo{author}{Liu, H.} \emph{et~al.}
\newblock \bibinfo{title}{Field test of twin-field quantum key distribution
  through sending-or-not-sending over 428 km}.
\newblock \emph{\bibinfo{journal}{Phys. Rev. Lett.}}
  \textbf{\bibinfo{volume}{126}}, \bibinfo{pages}{250502}
  (\bibinfo{year}{2021}).

\bibitem{chen21_511km_TFQKD}
\bibinfo{author}{Chen, J.-P.} \emph{et~al.}
\newblock \bibinfo{title}{Twin-field quantum key distribution over a 511~km
  optical fibre linking two distant metropolitan areas}.
\newblock \emph{\bibinfo{journal}{Nat. Photonics}}
  \textbf{\bibinfo{volume}{15}}, \bibinfo{pages}{570--575}
  (\bibinfo{year}{2021}).

\bibitem{clivati22}
\bibinfo{author}{Clivati, C.} \emph{et~al.}
\newblock \bibinfo{title}{Coherent phase transfer for real-world twin-field
  quantum key distribution}.
\newblock \emph{\bibinfo{journal}{Nat. Commun.}} \textbf{\bibinfo{volume}{13}},
  \bibinfo{pages}{157} (\bibinfo{year}{2022}).

\bibitem{wang22}
\bibinfo{author}{Wang, S.} \emph{et~al.}
\newblock \bibinfo{title}{Twin-field quantum key distribution over 830-km
  fibre}.
\newblock \emph{\bibinfo{journal}{Nat. Photonics}}
  \textbf{\bibinfo{volume}{16}}, \bibinfo{pages}{154 -- 161}
  (\bibinfo{year}{2022}).

\bibitem{Chen22_658km_TF-QKD}
\bibinfo{author}{Chen, J.-P.} \emph{et~al.}
\newblock \bibinfo{title}{Quantum key distribution over 658 km fiber with
  distributed vibration sensing}.
\newblock \emph{\bibinfo{journal}{Phys. Rev. Lett.}}
  \textbf{\bibinfo{volume}{128}}, \bibinfo{pages}{180502}
  (\bibinfo{year}{2022}).

\bibitem{toliver03}
\bibinfo{author}{Toliver, P.} \emph{et~al.}
\newblock \bibinfo{title}{Experimental investigation of quantum key
  distribution through transparent optical switch elements}.
\newblock \emph{\bibinfo{journal}{IEEE Photon. Technol. Lett.}}
  \textbf{\bibinfo{volume}{15}}, \bibinfo{pages}{1669--1671}
  (\bibinfo{year}{2003}).

\bibitem{tang16_MDI_Network}
\bibinfo{author}{Tang, Y.-L.} \emph{et~al.}
\newblock \bibinfo{title}{Measurement-device-independent quantum key
  distribution over untrustful metropolitan network}.
\newblock \emph{\bibinfo{journal}{Phys. Rev. X}} \textbf{\bibinfo{volume}{6}},
  \bibinfo{pages}{011024} (\bibinfo{year}{2016}).

\bibitem{zhong22_TF_Network}
\bibinfo{author}{Zhong, X.}, \bibinfo{author}{Wang, W.},
  \bibinfo{author}{Mandil, R.}, \bibinfo{author}{Lo, H.-K.} \&
  \bibinfo{author}{Qian, L.}
\newblock \bibinfo{title}{Simple multiuser twin-field quantum key distribution
  network}.
\newblock \emph{\bibinfo{journal}{Phys. Rev. Appl.}}
  \textbf{\bibinfo{volume}{17}}, \bibinfo{pages}{014025}
  (\bibinfo{year}{2022}).

\bibitem{Xie22_AsynchronousMDI-QKD}
\bibinfo{author}{Xie, Y.-M.} \emph{et~al.}
\newblock \bibinfo{title}{Breaking the rate-loss bound of quantum key
  distribution with asynchronous two-photon interference}.
\newblock \emph{\bibinfo{journal}{PRX Quantum}} \textbf{\bibinfo{volume}{3}},
  \bibinfo{pages}{020315} (\bibinfo{year}{2022}).

\bibitem{Zeng22_MP}
\bibinfo{author}{Zeng, P.}, \bibinfo{author}{Zhou, H.}, \bibinfo{author}{Wu,
  W.} \& \bibinfo{author}{Ma, X.}
\newblock \bibinfo{title}{Quantum key distribution surpassing the repeaterless
  rate-transmittance bound without global phase locking}
  (\bibinfo{year}{2022}).
\newblock \bibinfo{note}{ArXiv: 2201.04300}.

\bibitem{xu20_SNS_AOPP_original}
\bibinfo{author}{Xu, H.}, \bibinfo{author}{Yu, Z.-W.}, \bibinfo{author}{Jiang,
  C.}, \bibinfo{author}{Hu, X.-L.} \& \bibinfo{author}{Wang, X.-B.}
\newblock \bibinfo{title}{Sending-or-not-sending twin-field quantum key
  distribution: Breaking the direct transmission key rate}.
\newblock \emph{\bibinfo{journal}{Phys. Rev. A}}
  \textbf{\bibinfo{volume}{101}}, \bibinfo{pages}{042330}
  (\bibinfo{year}{2020}).

\bibitem{Jiang20_zigzag_AOPP}
\bibinfo{author}{Jiang, C.}, \bibinfo{author}{Hu, X.-L.}, \bibinfo{author}{Xu,
  H.}, \bibinfo{author}{Yu, Z.-W.} \& \bibinfo{author}{Wang, X.-B.}
\newblock \bibinfo{title}{Zigzag approach to higher key rate of
  sending-or-not-sending twin field quantum key distribution with finite-key
  effects}.
\newblock \emph{\bibinfo{journal}{New J. Phys.}} \textbf{\bibinfo{volume}{22}},
  \bibinfo{pages}{053048} (\bibinfo{year}{2020}).

\bibitem{hu19_asymmetric_SNS}
\bibinfo{author}{Hu, X.-L.}, \bibinfo{author}{Jiang, C.}, \bibinfo{author}{Yu,
  Z.-W.} \& \bibinfo{author}{Wang, X.-B.}
\newblock \bibinfo{title}{Sending-or-not-sending twin-field protocol for
  quantum key distribution with asymmetric source parameters}.
\newblock \emph{\bibinfo{journal}{Phys. Rev. A}}
  \textbf{\bibinfo{volume}{100}}, \bibinfo{pages}{062337}
  (\bibinfo{year}{2019}).

\bibitem{Maurice20}
\bibinfo{author}{Maurice, V.} \emph{et~al.}
\newblock \bibinfo{title}{Miniaturized optical frequency reference for
  next-generation portable optical clocks}.
\newblock \emph{\bibinfo{journal}{Opt. Express}} \textbf{\bibinfo{volume}{28}},
  \bibinfo{pages}{24708--24720} (\bibinfo{year}{2020}).

\bibitem{Talvard17}
\bibinfo{author}{Talvard, T.} \emph{et~al.}
\newblock \bibinfo{title}{Enhancement of the performance of a fiber-based
  frequency comb by referencing to an acetylene-stabilized fiber laser}.
\newblock \emph{\bibinfo{journal}{Opt. Express}} \textbf{\bibinfo{volume}{25}},
  \bibinfo{pages}{2259--2269} (\bibinfo{year}{2017}).

\bibitem{Caspar20}
\bibinfo{author}{Caspar, P.} \emph{et~al.}
\newblock \bibinfo{title}{Heralded distribution of single-photon path
  entanglement}.
\newblock \emph{\bibinfo{journal}{Phys. Rev. Lett.}}
  \textbf{\bibinfo{volume}{125}}, \bibinfo{pages}{110506}
  (\bibinfo{year}{2020}).

\bibitem{Pompili21}
\bibinfo{author}{Pompili, M.} \emph{et~al.}
\newblock \bibinfo{title}{Realization of a multinode quantum network of remote
  solid-state qubits}.
\newblock \emph{\bibinfo{journal}{Science}} \textbf{\bibinfo{volume}{372}},
  \bibinfo{pages}{259--264} (\bibinfo{year}{2021}).

\bibitem{Yu19_SNS_statistical}
\bibinfo{author}{Yu, Z.-W.}, \bibinfo{author}{Hu, X.-L.},
  \bibinfo{author}{Jiang, C.}, \bibinfo{author}{Xu, H.} \&
  \bibinfo{author}{Wang, X.-B.}
\newblock \bibinfo{title}{Sending-or-not-sending twin-field quantum key
  distribution in practice}.
\newblock \emph{\bibinfo{journal}{Sci.Rep.}} \textbf{\bibinfo{volume}{9}},
  \bibinfo{pages}{1--8} (\bibinfo{year}{2019}).

\bibitem{Jiang19_SNS_composable_security}
\bibinfo{author}{Jiang, C.}, \bibinfo{author}{Yu, Z.-W.}, \bibinfo{author}{Hu,
  X.-L.} \& \bibinfo{author}{Wang, X.-B.}
\newblock \bibinfo{title}{Unconditional security of sending or not sending
  twin-field quantum key distribution with finite pulses}.
\newblock \emph{\bibinfo{journal}{Phys. Rev. Appl.}}
  \textbf{\bibinfo{volume}{12}}, \bibinfo{pages}{024061}
  (\bibinfo{year}{2019}).

\end{thebibliography}
\end{document}


\title{Supplementary Information \\Twin-field quantum key distribution without optical frequency dissemination}

\author{Lai Zhou}
\author{Jinping Lin}
\author{Yumang Jing}
\author{Zhiliang Yuan}
\affiliation{Beijing Academy of Quantum Information Sciences, Beijing 100193, China}

\maketitle
\section{Encoder}
The purpose of the encoder is to turn the continuous-wave input into a 1~GHz, 300~ps pulse train with each pulse's intensity and phase set according to the requirements by a TF-QKD protocol. The encoder is capable of supporting all TF-QKD protocols, although just the SNS protocol was demonstrated in this work.
Below, we describe the encoder with reference to Fig.~2\textbf{b}, Main Text.

The input light has its polarisation aligned to the slow axis of a series of intensity and phase modulators (3 IM's and 2 PM's), which are driven by a 25~GSa/s arbitrary waveform generator containing 2 waveform and 4 pattern output channels.
IM$_1$ is modulated for pulse carving, IM$_2$ by a waveform signal for preparing pulses of different intensities, and IM$_3$ by a second binary signal for adjusting the intensity contrast between the `quantum reference' and the quantum signal.  Together, three IM's prepare pulses of five different intensity levels: $\mu_{qr}$ (quantum reference), $\mu_Z$ (signal state), $\mu_2$ (strong decoy), $\mu_1$ (weak decoy) and $\mu_0$ (vacuum).  The encoder is able to achieve $>$40~dB extinction ratio between the signal ($\mu_Z$) and vacuum ($\mu_0$) states.
After intensity modulation, PM$_1$ and PM$_2$ encode the phase of each quantum signal pulse with one of 16 phase values, $\theta \in \{0,\pi/8,2\pi/8...15\pi/8\}$ meeting TF-QKD's phase randomisation and qubit encoding requirement, while the quantum reference pulses are left unmodulated.   The quantum signals are interleaved with the quantum reference in every 100~ns. Overall, the quantum signals occupy 50~\% of the total time slots and  have an effective clock rate of 500~MHz. In our experiments, we used a pattern length of 40000 bits, corresponding to a duration of 40~$\mu$s.  Modulation patterns were carefully designed to produce correct sending and matching probabilities of different class of pulses as expected from the TF-QKD protocol.

All optical elements in each encoder are installed in an enclosure for stability.

\section{Active Feedbacks}

\begin{figure*}[hbt]
\includegraphics[width=1.5\columnwidth]{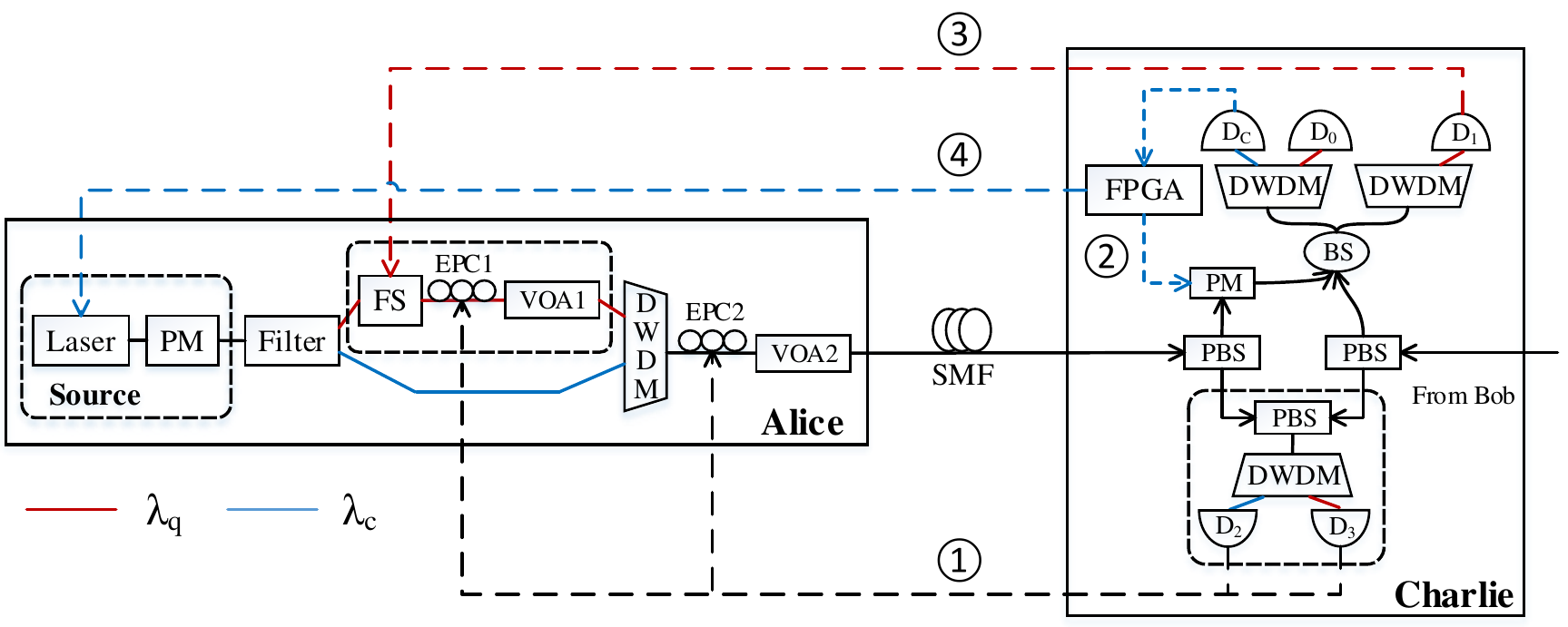}
\caption{\textbf{Active feedbacks.} Four feedback routines are implemented. \protect\circled{1} Polarisation precompensation using detectors D$_2$ and D$_3$ as error signals to adjust EPC2 and EPC1; Rate: 5 -- 10~Hz. \protect\circled{2}  Fast phase compensation for the channel reference $\lambda_c$ using detector D$_c$ as error signal to drive Charlie's phase modulator (PM); Rate: 200~kHz; \protect\circled{3}  Active phase compensation for the quantum wavelength $\lambda_q$ using detector  D$_1$'s count rate as error signal to adjust Alice's fibre stretcher (FS); Rate: 50 -- 100~Hz; \protect\circled{4}  Active correction of the laser frequency difference. Rate: daily.  Routine \protect\circled{1} applies on both Alice and Bob,  \protect\circled{2} locally at Charlie, \protect\circled{3}  and \protect\circled{4}  apply just on Alice.
BS: beam splitter; DWDM: dense wavelength division multiplexer;  EPC: electronically driven polarisation controller;   PBS: polarisation beam splitter/combiner; PM: phase modulator; SMF: single mode fibre; VOA: variable optical attenuator. }
\label{fig:feedback}
\end{figure*}

To ensure stable operation, we implemented four feedback routines to correct the photon polarisation, the fast phase drift, the slow residual phase drift, and the laser frequency difference.  These routines require each user to have two mutually coherent wavelengths of $\lambda_c$ and $\lambda_q$ for channel reference and quantum reference/signal.

As shown in Fig.~\ref{fig:feedback}, each user's $\lambda_q$ and $\lambda_c$ signals of 100~GHz spacing are derived from the same laser via electro-optic frequency comb generation.
They are separated by an optical filter with 55~dB isolation into two separate paths. The encoding path contains a fibre stretcher (FS),  an electronically driven polarisation controller (EPC1) and a variable optical attenuator (VOA1) in addition to the modulation elements (not shown).  The channel reference path is just a straight fibre.  The two paths are multiplexed together by a 50 GHz Dense Wavelength Division Multiplexer (DWDM) into a second EPC (EPC2) and then sent to Charlie via the quantum link segment after being attenuated to the desired intensity by a second optical attenuator (VOA2).  VOA1 sets the intensity contrast to a desired level between the two wavelengths, while VOA2 is to set the quantum signals to the correct fluxes as defined in Table~\ref{tab:parameters}.

At each input, Charlie uses a polarisation beam splitter (PBS) to ensure all photons entering his 50/50 interfering beam splitter (BS) to have an identical polarisation and thus achieve maximum interference visibility.   Polarisation rotation by the quantum link segment will affect the transmission through the PBS. The reflected signals by the PBS is routed to two single photon detectors (D$_2$ and D$_3$) via a polarisation beam combiner and a DWDM.  D$_2$ (D$_3$) detects the $\lambda_c$ ($\lambda_q$) photons and its count rate is minimised via controlling EPC2 (EPC1), thus maximising the transmissions.  This feedback routine operates continuously at a rate of 5 - 10~Hz.

The interference outcome between Alice and Bob's channel references ($\lambda_c$) is detected by D$_c$ and its count rate feeds to the FPGA PID controller which then computes a compensation voltage to drive Charlie's PM at a feedback rate of 200~kHz.   This feedback locks Alice and Bob's channel reference signals to a differential phase of $\pi/2$, and corrects Alice and Bob's optical frequency difference and the fibre fluctuation of the entire quantum channel.
The  mutual coherence ensures the differential phase of the $\lambda_q$ wavelength is mostly compensated as well, with its residual amounting to about $|\lambda_q-\lambda_c|/\lambda_q$ of its original rate.
This residual drift is then corrected for by Alice's FS with the feedback signal from D$_1$'s count rate.  This slow feedback operates at a rate of 50 -- 100~Hz.

As described in Main Text (see Fig.~3\textbf{c}), the FPGA PID controller allows instantaneous readout of the laser frequency difference even when the lasers are separated by hundreds of kilometers fibres.   Once the frequency offset exceeds a certain value, \textit{e.g.}, 300~Hz,  Charlie transmits an instruction to Alice to offset her laser frequency accordingly. This feedback is applied less than daily thanks to the stability of the lasers.

\section{System loss characterisation}

In our setup, the quantum channel is made of ultra-low-loss fibre spools (Corning G654.C ULL) of mainly 25.2~km and 50.4~km each in length.  They were spliced into sets of different lengths to allow varying fibre lengths in experiments. Including loss from fibre connectors, the quantum channel was characterised to have an average loss coefficient of 0.168~dB~km$^{-1}$.
Table~\ref{tab:fiber_loss} summarises the lengths and their corresponding losses for the fibre configurations used in our experiments.

Inside Charlie's module, there is a chain of fibre components from either user's input port to the single photon detectors.  These components and their losses are summarised in Table~\ref{tab:Charlie_loss}.
Charlie's transmission loss at Alice's side is 4.5 dB, which is 2~dB higher  than Bob's side because of the extra phase modulator.
This 2~dB asymmetry was compensated either by a 2~dB attenuator in the symmetric 403.73~km experiment or by a 10~km fibre spool for all other experiments.

We present the performance of superconducting nanowire single photon detectors (SNSPD's: D$_0$ and D$_1$) in Table~\ref{tab:detectors}.  These SNSPDs are polarisation-sensitive, so a manual polarisation controller is placed in front of each SNSPD for maximising the efficiency. Their noise counts include both detector dark counts (2~Hz) and scattered noise (2~Hz) from the fibre channel.

\begin{table*}[hbt]
\caption{Lengths and corresponding losses for the fibre links we used in the two sets of experiments.}
\centering
\begin{tabular}{c|c c|c c|c c}
\hline
\hline
\multirow{2}{*}{}& \multicolumn{2}{c}{Total} & \multicolumn{2}{|c}{Alice} &  \multicolumn{2}{|c}{Bob}\\
\cline{2-7}
 &length (km) & loss (dB) & length (km) & loss (dB) & length (km) & loss (dB)\\
\hline
\hline
\multirow{3}{*}{Symmetric}
 & 403.73  & 67.89 & 201.87 & 33.97 & 201.86 & 33.92\\
& 518.16 & 87.02 & 254.38 & 42.57 & 263.78 & 44.45\\
& 615.59 & 103.27 & 302.80 & 50.55 & 312.79 & 52.72\\
\hline
\multirow{2}{*}{Asymmetric}
& 455.65 & 76.68 & 201.87 & 34.16 & 253.78 & 42.52 \\
& 407.23 & 68.55 & 153.45 & 26.03 & 253.78 & 42.52\\
\hline
\hline
\end{tabular}
\label{tab:fiber_loss}
\end{table*}

\begin{table}[hbt]
\caption{Charlie's components loss.}
\centering
\begin{tabular}{c|p{1.2cm}|p{1.2cm}}
\hline
\hline
 & Alice & Bob\\
\hline\hline
Polarisation beam splitter  & 0.7 & 0.7\\
\hline
Phase modulator & 2.0  &  n/a\\
\hline
50/50 beam splitter  & 0.3 & 0.3 \\
\hline
DWDM filter & 1.2 & 1.2\\
\hline
Polarisation controller & 0.3 & 0.3\\
\hline
Total loss (dB) & 4.5 & 2.5\\
\hline\hline
\end{tabular}
\label{tab:Charlie_loss}
\end{table}

\begin{table}[hbt]
\caption{Performance of Charlie's detectors D$_0$ and D$_1$.}
\centering
\begin{tabular}{c|c|c}
\hline
\hline
 Detector & Efficiency & Noise Count Rate \\
 \hline\hline
D$_0$ & 60~\% & 4 Hz \\
\hline
D$_1$ & 65~\%  & 4 Hz  \\
\hline
\hline
\end{tabular}
\label{tab:detectors}
\end{table}

\section{Visibility characterisation}

In this section, we describe how we measured the various interference visibilities reported in the Main Text.

To measure the interference visibility over a free-drifting long quantum channel, we use an optical power meter with a sampling rate of 500~kHz.  Each measurement has a duration of 10~s and produces a corresponding data array of $5\times10^6$ samples.  We compute a visibility ($V$) for each segment of $2 \times 10^4$ samples (40~ms) from the average ($\bar{I}_{max}$) of the 5 highest values and the average  ($\bar{I}_{min}$) of the 5 lowest values,
\begin{equation}
    V = \frac{\bar{I}_{max} - \bar{I}_{min}}{\bar{I}_{max} - \bar{I}_{min}}.
\label{eq:visibility}
\end{equation}
\noindent We then calculate the average visibility from the entire 10~s data and its standard deviation.

\begin{figure}[ht]
\includegraphics[width=1\columnwidth]{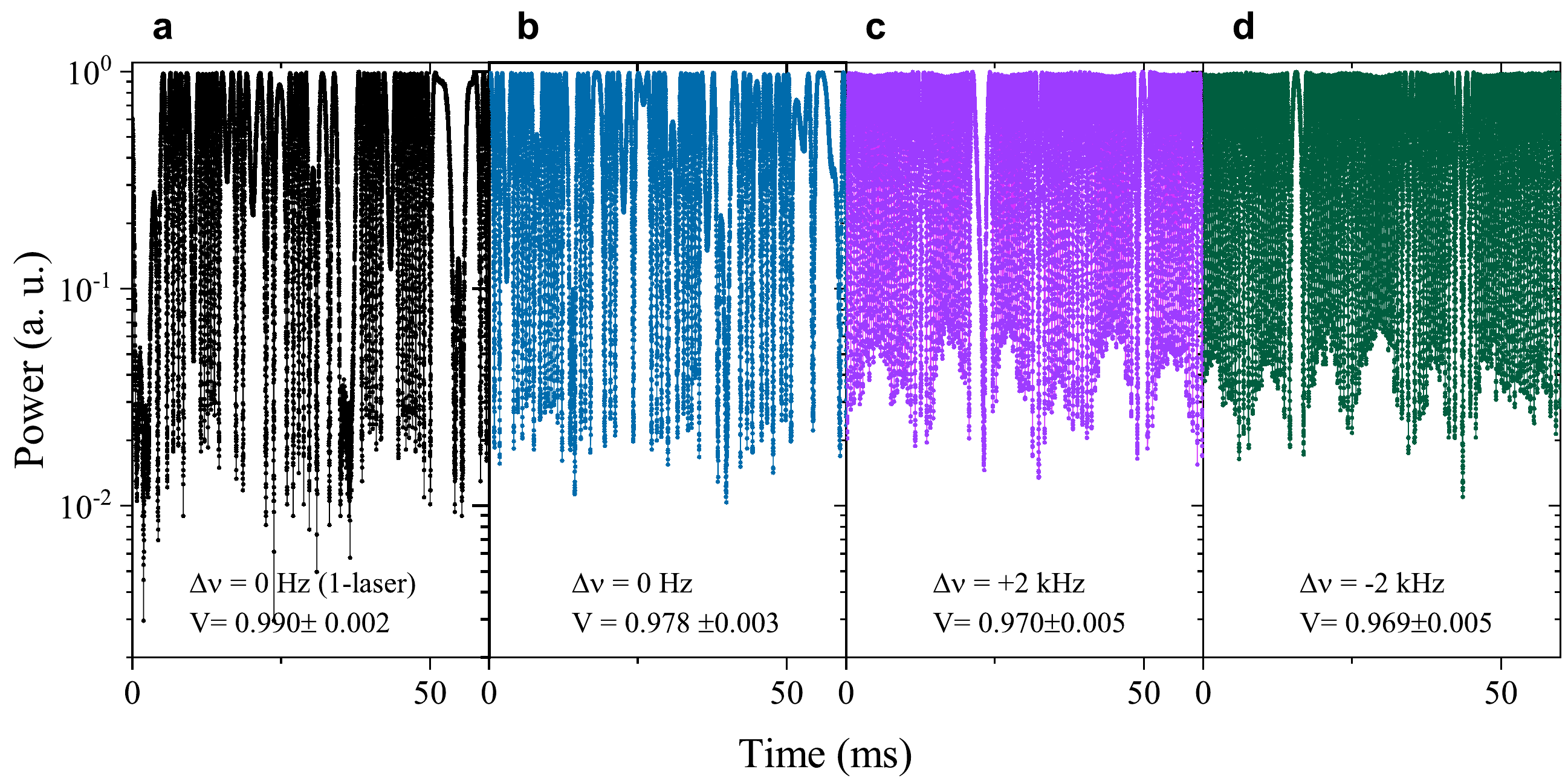}
\caption{\textbf{Free-drifting interference fringes.}  All data were measured with the 615.6~km (302.8~km + 312.8~km) that drifts freely and a power meter at a sampling rate of 500~kHz. \textbf{a}, Alice and Bob shared a common laser; \textbf{b}, Independent lasers with nominally 0-Hz frequency offet; \textbf{c}, As \textbf{b} but with +2~kHz frequency offset; \textbf{d}, As \textbf{b} but with -2~kHz offset.}
\label{fig:fast_fringes}
\end{figure}

Fig.~\ref{fig:fast_fringes} shows the measured interference fringes for the channel reference over the free-drifting quantum channel of 615.6~km.  When Alice and Bob shared a common laser (\textbf{a}), the measured visibility is just 99~\% and this imperfect visibility is caused mainly by the 10~km asymmetry in the quantum link.
The visibility drops to 97.8~\% when Alice and Bob use truly independent lasers with nominally 0~Hz offset (\textbf{b}), illustrating the performance penalty by the loss of mutual frequency locking.
Increasing the offset to $2$~kHz (\textbf{c}, \textbf{d}), the measured visibility further decreases to $\sim$97.0 \%. However, this further decrease is attributed mainly to the limitation of the power meter sampling rate, because the absolute drift rate is higher in the presence of a frequency offset, as shown in Fig.~3\textbf{b} of the Main Text.
Note that Figs.~\ref{fig:fast_fringes}\textbf{c} and \textbf{d} can also have destructive interference occasionally as low as those lowest in \textbf{b} when the net drift rate happens to be slow.

\begin{figure}[ht]
\includegraphics[width=0.8\columnwidth]{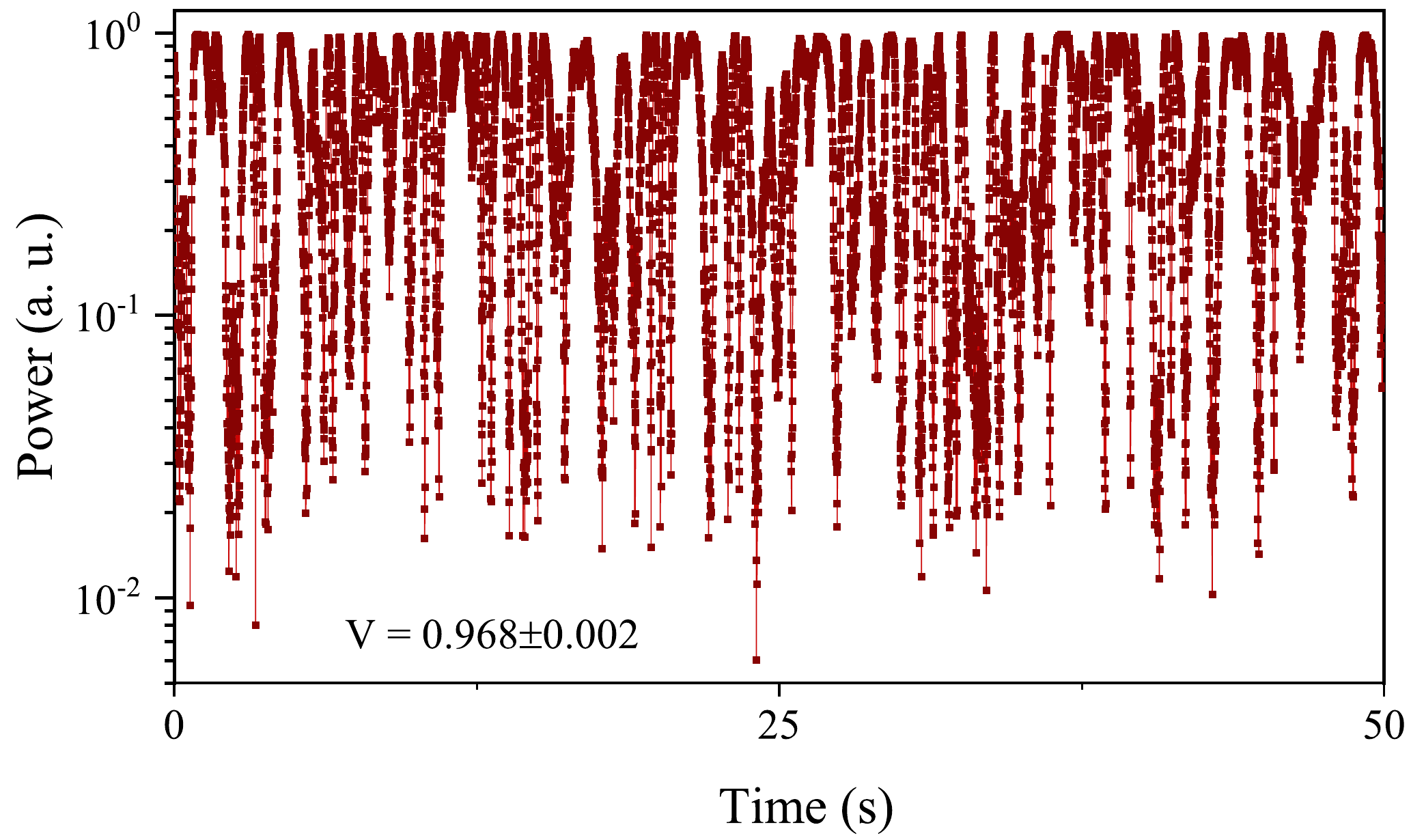}
\caption{\textbf{Interference fringes of the quantum wavelength.}  The 615.6~km (302.8~km + 312.8~km) was stabilised by the FPGA PID controller and  the data were recorded by
a power meter at 200~Hz.  Alice and Bob used independent lasers with nominally 0~Hz detuning. }
\label{fig:slow_fringes}
\end{figure}

Once the quantum channel is stabilised by the FPGA PID controller, the residual phase drift of the quantum signal becomes much slower.  In this case, we use a power meter at a slower sampling rate of 200~Hz. We record a 50~s data containing $10^4$ samples. From every 10~s of data (2000 samples),  we compute a visibility value from averaging 10 highest and 10 lowest values, and then compute the average visibility and its standard deviation.   Fig.~\ref{fig:slow_fringes} reports the measurement result for the 615.6~km quantum channel  with two independent lasers.  The average visibility is 96.8~\%.

\section{Detailed experimental parameters and  results}

In Table~\ref{tab:parameters}, we give the detailed experimental encoding parameters used for symmetrical and asymmetrical setups. Table~\ref{tab:symmetric} and \ref{tab:asymmetric} summarize the experimental results and quantities relative to the key rate calculation using AOPP for symmetric and asymmetric cases, respectively. The total number of signal pulses sent is given by $N_{\text{tot}}$. The number of valid detections reported by Charlie is denoted as ``Detected $\text{AB}_{\text{ab}}$", where ``A'' (``B'') is X or Z indicating the basis Alice (Bob) chooses; ``a'' (``b'') is 0, 1, 2 or 3 indicating the intensity Alice (Bob) chooses is $\mu_{A0}$ ($\mu_{B0}$), $\mu_{A1}$ ($\mu_{B1}$), $\mu_{A2}$ ($\mu_{B2}$), or $\mu_{Az}$ ($\mu_{Bz}$), respectively. ``QBER ($\text{X}_{11}$)'' and ``QBER ($\text{X}_{22}$)'' represent the error rates in X bases for the decoy states with respective intensity ``11'' and ``22''. The rest of the notations is explained in the main text.

\begin{table*}[ht]
\caption{Encoding parameters used in the two sets of experiments. In symmetric experiments, Alice and Bob share an identical set of parameters.   $\mu_Z$, $\mu_2$, $\mu_1$ and $\mu_0$ are the values of mean photon number per pulse for  signal, strong decoy, weak decoy and vacuum pulses. $P_Z$ is the probability of a pulse to be encoded in the coding (Z) basis. The parameter $\epsilon$ is the probability of a signal pulse to be actually sent when Z basis is chosen. $P_X = 1 - P_Z$  is the selection probability of the checking (X) basis, while $p_{\mu_2}$, $p_{\mu_1}$  and $p_{\mu_0}$ are the selection probabilities for preparing $\mu_2$, $\mu_1$ and $\mu_0$ decoy states in the X-encoding basis. }
\centering
\begin{tabular}{c|c|c|c}
\hline\hline
\multirow{3}{*}{Parameter}{} & Symmetric & \multicolumn{2}{|c}{ \quad \quad\quad Asymmetric \quad \quad\quad \quad}  \\
& 403.7km/518.2km/615.6km & \multicolumn{2}{|c}{ 407.2km/455.7km }\\
\cline{2-4}
& Alice/Bob & \quad \quad  Alice \quad \quad&  Bob  \\
\hline
$\mu_Z$ & 0.493 &  0.493 & 0.493 \\
$\mu_2$ & 0.493 & 0.114 & 0.493 \\
$\mu_1$ & 0.105 & 0.057 & 0.246 \\
$\mu_0$ & 0.0002 & 0.0002 & 0.0002 \\
\hline
$P_Z$ & 0.735 & ~~0.735~~ & 0.735 \\
$P_X$ 
& 0.265 & 0.265 & 0.265\\
$\epsilon$ & 0.269 & 0.136 & 0.405\\
$p_{\mu_2}$ & 0.216 & 0.216 & 0.216\\
$p_{\mu_1}$ & 0.706 & 0.706 & 0.706\\
$p_{\mu_0}$ 
& 0.078 & 0.078 & 0.078 \\
\hline\hline
\end{tabular}
\label{tab:parameters}
\end{table*}

\begin{table}[ht]
\centering
\caption{Finite-size symmetric SNS with AOPP: experimental results at various quantum link fibre lengths. }
\scalebox{0.8}{
\begin{tabular}{cccc}
\toprule
Total length (km) & 403.73 & 518.16 & 615.59  \\
Alice - Charlie (km) & 201.87 & 254.38 & 302.80 \\
Bob - Charlie (km) & 201.86 & 263.78 & 312.79 \\
\midrule
$N_{\text{tot}}$ & \quad $2.025 \times 10^{12}$ \quad & \quad $2.475 \times 10^{12}$ \quad & \quad $1.4175 \times 10^{13}$ \\
Number of phase slices & 16 & 16 &16 \\
\midrule
Detected $\text{XX}_{20}$ & 93574 & 15168 & 13360 \\
Detected $\text{XX}_{02}$ & 102956 & 11292 & 18185 \\
Detected $\text{XX}_{10}$ & 60774 & 11428 & 11201 \\
Detected $\text{XX}_{01}$ & 63251 & 12807 &  12698 \\
Detected $\text{XX}_{00}$ & 38 & 14 &   85 \\
Detected $\text{XZ}_{00}$ & 814 & 358 &  1464 \\
Detected $\text{XZ}_{10}$ & 1656903 & 296696 &  310666 \\
Detected $\text{XZ}_{20}$ & 2336341 & 418785 &  402898 \\
Detected $\text{ZX}_{00}$ & 788 & 389 &  1624 \\
Detected $\text{ZX}_{01}$ & 1650969 & 293079 &  323169 \\
Detected $\text{ZX}_{02}$ & 2333340 & 418695 &   419565\\
Detected $\text{ZZ}_{03}$ & 8020164 & 1420140 &  1492968 \\
Detected $\text{ZZ}_{30}$ & 8390554 & 1440863 &  1400694 \\
Detected $\text{ZZ}_{33}$ & 6121571 & 1082752 &  1050251 \\
Detected $\text{ZZ}_{00}$ & 21033 & 9178 &  40051 \\
\midrule
QBER ($\text{X}_{11})$  & 5.17\% & 4.81\% & 4.75\% \\
QBER ($\text{X}_{22})$  & 4.77\% & 5.02\% & 5.12\% \\
\midrule
QBER ($E_z$ before AOPP) & 27.24\% &  27.62\% & 27.37\% \\
QBER ($E_z$ after AOPP) & 0.19\% & 0.48\% & 1.97\% \\
$n_1$ (Before AOPP) & $9.565 \times 10^6$ & $1.665 \times 10^6$ & $1.744 \times 10^6$ \\
$n_1$ (After AOPP) & $1.616 \times 10^6$ & $2.763 \times 10^5$ & $2.984 \times 10^5$\\
$e_1^{ph}$ (Before AOPP) & 7.09\% & 7.57\% & 6.29\%\\
$e_1^{ph}$ (After AOPP) & 13.36\% & 14.49\% & 12.24\%\\
\midrule
SKR (bit/s) & 146.70 & 14.38 & 0.32 \\
SKR (bit/signal) & $2.934 \times 10^{-7}$ & $2.875 \times 10^{-8}$& $6.366 \times 10^{-10}$\\
$\text{SKC}_0$ (bit/signal) & $2.380 \times 10^{-7}$ &  $2.845 \times 10^{-9}$ & $6.565 \times 10^{-11}$ \\
Ratio SKR over $\text{SKC}_0$ & 1.23 & 10.11 & 9.70 \\

\bottomrule

\end{tabular}
}
\label{tab:symmetric}
\end{table}
\begin{table}[ht]

\centering
\caption{Finite-size asymmetric SNS with AOPP: experimental results at various quantum link fibre lengths.}
\scalebox{0.8}{
\begin{tabular}{ccc}
\toprule
Total length (km) & 407.23 & 455.65   \\
Alice - Charlie (km) & 153.45 & 201.87   \\
Bob - Charlie (km) & 253.78 & 253.78 \\
\midrule
$N_{\text{tot}}$ & \quad $2.25 \times 10^{12}$ \quad & \quad $2.2 \times 10^{12}$ \quad  \\
Number of phase slices & 16 & 16  \\
\midrule
Detected $\text{XX}_{20}$ & 20887 & 27708  \\
Detected $\text{XX}_{02}$ & 19018 &  20497 \\
Detected $\text{XX}_{10}$ & 37937 & 36554  \\
Detected $\text{XX}_{01}$ & 36669 & 33340  \\
Detected $\text{XX}_{00}$ & 20 & 27  \\
Detected $\text{XZ}_{00}$ & 369 & 453  \\
Detected $\text{XZ}_{10}$ & 785947 & 804659  \\
Detected $\text{XZ}_{20}$ & 474298 & 492710  \\
Detected $\text{ZX}_{00}$ & 610 & 558  \\
Detected $\text{ZX}_{01}$ & 1118620 & 1192466  \\
Detected $\text{ZX}_{02}$ & 669280 & 699462 \\
Detected $\text{ZZ}_{03}$ & 3499529 & 3672181  \\
Detected $\text{ZZ}_{30}$ & 3611502 & 3679227  \\
Detected $\text{ZZ}_{33}$ & 2956649 & 2999841  \\
Detected $\text{ZZ}_{00}$ & 10955 & 12133  \\
\midrule
QBER ($\text{X}_{11})$  & 5.31\% & 5.07\%  \\
QBER ($\text{X}_{22})$  & 5.10\% & 5.19\%  \\
\midrule
QBER ($E_z$ before AOPP) & 29.44\% &  29.06\%  \\
QBER ($E_z$ after AOPP) & 0.26\% & 0.27\%  \\
$n_1$ (Before AOPP) & $4.251 \times 10^6$ & $4.354 \times 10^6$  \\
$n_1$ (After AOPP) & $6.963 \times 10^5$ & $7.084 \times 10^5$ \\
$e_1^{ph}$ (Before AOPP) & 8.21\% & 7.74\% \\
$e_1^{ph}$ (After AOPP) & 15.43\% & 14.60\% \\
\midrule
SKR (bit/s) & 46.31 & 50.75 \\
SKR (bit/signal) & $9.261 \times 10^{-8}$ & $1.015 \times 10^{-7}$\\
$\text{SKC}_0$ (bit/signal) & $2.078 \times 10^{-7}$ &  $3.193\times 10^{-8}$  \\
Ratio SKR over $\text{SKC}_0$ & 0.45 & 3.18  \\
\bottomrule

\end{tabular}
}
\label{tab:asymmetric}
\end{table}
